\documentclass[pra,aps,superscriptaddress,amsmath,amssymb,twocolumn,floatfix]{revtex4-1}
\pdfoutput=1
\usepackage[pdftex]{graphicx}
\usepackage{bm}
\usepackage{hyperref}
\usepackage[usenames,dvipsnames]{color}

\def\mathbold{\bf}

\newcommand{\bg}{ \begin{gather} }
\newcommand{\eg}{\end{gather}}

\def\br{{\mathbold r}}

\def\eps{\epsilon}

\newcommand{\corr}[1]{\langle #1\rangle}
\newcommand{\Tr}{\mathop{\rm Tr}}

\newcommand{\tr}{\mathop{\rm tr}}

\newcommand{\const}{\text{const}}
\renewcommand{\Re}{\mathop{\rm Re}}
\renewcommand{\Im}{\mathop{\rm Im}}

\def\be{\begin{equation}}
\def\ee{\end{equation}}

\begin{document}

\title{Superconducting fluctuations at arbitrary disorder strength}

\author{Nikolai A. Stepanov}
\affiliation{Skolkovo Institute of Science and Technology, Skolkovo, Moscow Region, 143026, Russia}
\affiliation{L. D. Landau Institute for Theoretical Physics, Chernogolovka, Moscow Region, 142432, Russia}
\affiliation{National Research University Higher School of Economics, Moscow, 101000, Russia}
 
\author{Mikhail A. Skvortsov}
\affiliation{Skolkovo Institute of Science and Technology, Skolkovo, Moscow Region, 143026, Russia}
\affiliation{L. D. Landau Institute for Theoretical Physics, Chernogolovka, Moscow Region, 142432, Russia}

\date{\today}

\begin{abstract}
We study the effect of superconducting fluctuations on the conductivity of metals at arbitrary temperatures $T$ and impurity scattering rates $\tau^{-1}$. 
Using the standard diagrammatic technique but in the Keldysh representation, we derive the general expression for the fluctuation correction to the dc conductivity applicable for any space dimensionality and analyze it in the case of the film geometry. 
We observe that the usual classification in terms of the Aslamazov-Larkin, Maki-Thompson and density-of-states diagrams is to some extent artificial since these contributions produce similar terms, which partially cancel each other.
In the diffusive limit, our results fully coincide with recent calculations in the Keldysh technique.
In the ballistic limit near the transition, we demonstrate the absence of a divergent term $(T\tau)^2$ attributed previously to the density-of-states contribution.
In the ballistic limit far above the transition, the temperature-dependent part of the conductivity correction is shown to grow as $T\tau/\ln(T/T_c)$, where $T_c$ is the critical temperature.
\end{abstract}


\maketitle

\section{Introduction}

Superconducting transition is a second-order phase transition, with pairing correlations emerging in a continuous fashion when the temperature $T$ passes through the critical temperature $T_c$. In the mean-field description, Cooper pairs appear only below $T_c$, since their formation is energetically unfavorable above the transition. Nevertheless they can emerge as virtual excitations also for $T>T_c$, due to thermal fluctuations. Fluctuation formation of Cooper pairs in the normal state is known as superconducting fluctuations. They manifest themselves in various physical properties, which acquire a temperature-dependent correction growing at $T\to T_c$ \cite{Levanyuk,Ginzburg}.

\begin{figure}[b]
\includegraphics[width=0.96\columnwidth]{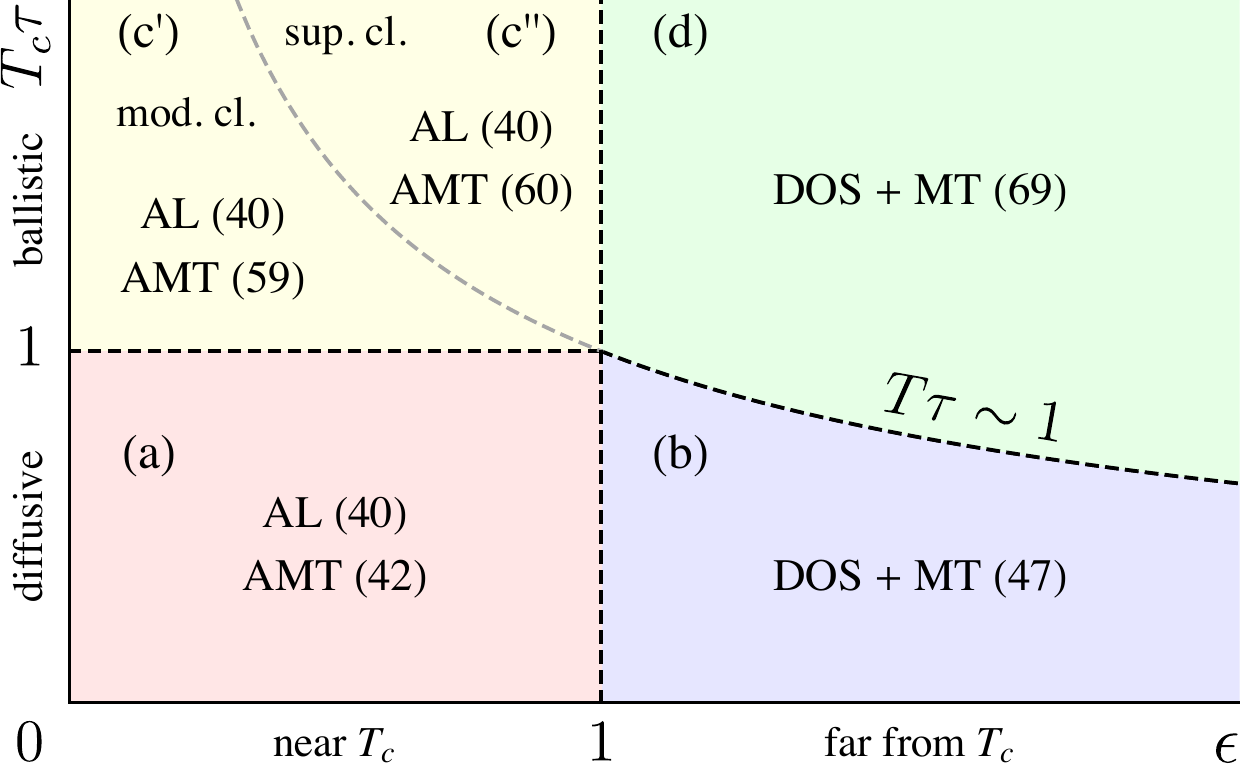}\hspace{0.02\columnwidth}
\caption{A sketch of the diagram for the fluctuation conductivity in the variables $\epsilon=\ln T/T_c$ and $T_c\tau$ (inverse disorder strength) showing the leading contributions and the corresponding formulae. The regions (a)--(d) label four asymptotic cases: close to $T_c$/far from $T_c$, and the diffusive/ballistic limits. 
The behavior of the anomalous MT (AMT) contribution near $T_c$ is different in the moderately clean (c$'$) and superclean (c$''$) regions separated by the line $T_c\tau\sqrt\epsilon\sim1$.
}\label{fig:phasediagram}
\end{figure}

The study of the effect of superconducting fluctuations on transport properties began with observation of fluctuation corrections to conductivity~\cite{Glover}. Theoretical understanding of the effect was elaborated in the seminal paper by Aslamazov and Larkin \cite{AL}, who calculated the direct contribution of fluctuation Cooper pairs to charge transport. Later on, an additional mechanism was analyzed by Maki \cite{MT1} and Thompson \cite{MT2}. Eventually it was recognized that there are three contributions to the fluctuation correction: paraconductivity (Aslamazov-Larkin, AL), scattering on superconducting fluctuations (Maki-Thompson, MT), and the contribution due to the depletion of the normal density of states (DOS) \cite{LV-book}. 
The theory developed in the beginning of 1970-ies was limited to the immediate vicinity of the critical temperature and described (apart from the AL term) the dirty limit, $T\tau\ll1$, where $\tau$ is the mean-free time
(we use the system of units with $e=\hbar=k_B=1$).
This limit is shown as region (a) in Fig.\ \ref{fig:phasediagram}.

In 1980-ies, the theory of fluctuation conductivity was extended to large temperatures above $T_c$ \cite{AV1980,Altshuler-Varlamov}, where the dirty limit was considered [region (b)]. The rise of high-temperature superconductivity in 1990-ies stimulated interest in less disordered superconductors, and the theory was generalized to the clean limit, $T\tau\gg1$, in the vicinity of $T_c$ \cite{AHL1995,Randeria,Dorin93,Livanov2000} [regions (a) and (c)]. The case of a superconductor without impurities was considered in Ref.\ \cite{Reggiani1991} [regions (c) and (d)].

In the presence of a magnetic field $H$, the superconducting transition takes place at the line $H_{c2}(T)$, above which superconducting fluctuations modify the normal-state conductivity. They have been studied in Ref.~\cite{GL2001}, where the low-temperature region in the vicinity of the quantum phase transition, $H\to H_{c2}$, has been analyzed in the diffusive limit. The general phase diagram in the plane ($H,T$) was addressed in Refs.\ \cite{Glatz2011,Tikhonov}. Frequency dependence of the conductivity correction for arbitrary $T>T_c$ in the diffusive limit was studied in Ref.\ \cite{Petkovic}.
Besides conductivity, superconducting fluctuations affect the Nernst coefficient \cite{Xu,Pourret,Ussishkin,SSVG,Michaeli}, that has a profound effect due to its smallness in the normal state.

Though the physical picture of superconducting fluctuations is rather well understood, their analytical description is a sufficiently complicated task. Despite half a century of studies, the dependence of the fluctuation conductivity on varius parameters is still debated: two recent papers \cite{Glatz2011,Tikhonov} predict a similar but different expressions for $\delta\sigma(H,T)$ in the dirty limit.
Historically, fluctuation corrections were calculated using the diagrammatic technique in the Matsubara representation, with the subsequent analytic continuation to real frequencies \cite{LV-book}. This method was employed also in Ref.\ \cite{Glatz2011}.
On the other hand, recent publications use a different Keldysh technique: either in the form of the Usadel equation \cite{Tikhonov} or in the sigma-model formalism \cite{Petkovic}. That makes comparison between different approaches problematic. 
Reported inconsistencies appeal for a revision of previously obtained results 
in the whole phase diagram of Fig.\ \ref{fig:phasediagram}.

Motivated by existing discrepancies, in this paper we derive a general expression for the dc conductivity correction for an arbitrary temperature, $T>T_c$, and disorder strength measured by the parameter $T\tau$, assuming an arbitrary space dimensionality $d$.
The final analysis will be performed in the experimentally relevant two-dimensional (2D) case corresponding to the film geometry. The existence of the exact expression interpolating between the four limiting cases discussed above provides us a tool for a critical review of previous results.

Our approach is based on the diagrammatic technique in the Keldysh representation.
The use of the diagrammatic language makes a bridge with the majority of previous calculations and allows us to compare individual contributions. On the other hand, the use of the Keldysh technique is crucial for accurate disorder averaging beyond the diffusive limit. Otherwise the number of terms with different analytic structure makes the routine procedure of analytic continuation extremely sophisticated.

The paper is organized as follows.
In Sec.\ \ref{S:General} we formulate the model, discuss the main ingredients of the technique and present the general expression for the conductivity correction.
In Secs.\ \ref{S:Diffusive-regime} and \ref{S:Ballistic-regime} we analyze dc conductivity in the diffusive and ballistic regimes, respectively. The crossover between the diffusive and ballistic regimes in the vicinity of the transition is studied in Sec.\ \ref{S:cross}.
In Sec.\ \ref{S:Discussion} we summarize our results and discuss them in the context of previous approaches. Finally, numerous technical details are relegated to several Appendices.

\section{General expression for the conductivity correction}
\label{S:General}

\subsection{Model}

We consider a disordered $s$-wave BCS superconductor described by the second-quantized Hamiltonian
\begin{equation}
\label{H}
H 
= 
\int d\br \biggl[ 
\sum_{\alpha} \psi_\alpha^{\dagger} \hat H_0 \psi_\alpha
- \frac{\lambda}{\nu}
\psi^\dagger_{\uparrow}\psi_{\downarrow}^\dagger
\psi_{\downarrow}\psi_{\uparrow}
\biggr] ,
\end{equation}
where $\psi_\alpha(\mathbf r)$ is the fermionic field, with $\alpha={\uparrow,\downarrow}$ labeling the spin degrees of freedom, $\lambda\ll1$ is the dimensionless interaction strength, and $\nu$ is the density of states at the Fermi energy (per one spin projection).

The single-particle Hamiltonian has the form:
\begin{equation}
\label{H0}
\hat H_0 = -\frac{(\nabla-i\mathbf{a})^2}{2m} - \mu + U(\br),
\end{equation}
where $\mu$ is the chemical potential.
We will consider the system at zero magnetic field, and the vector potential $\mathbf{a}$ will play the role of a source field to generate current and will be set to zero afterwards.
The disorder potential $U(\br)$ is assumed to be a Gaussian white noise specified by the correlation function
\be
\label{<UU>}
\corr{U(\br)U(\br')} 
= 
\frac{\delta(\br-\br')}{2\pi\nu\tau},
\ee
where $\tau$ is the mean-free time.

We assume that disorder is weak, with the quasiclassical parameter $\mu\tau\gg1$. This inequality allows us to apply the usual diagrammatic cross technique for disorder averaging. We also assume that the system is far from the metal-insulator transition, such that the weak-localization and interaction \cite{AA} corrections can be neglected.

\subsection{Keldysh technique}

In order to calculate the conductivity $\sigma$, we use the Keldysh technique for superconducting systems \cite{Kamenev,FLS2000,AKamenev}. Integrating out fermions, we obtain a theory formulated in terms of the classical, $\Delta_\text{cl}$, and quantum, $\Delta_\text{q}$, components of the order parameter and defined by the partition function
\begin{equation}
Z[\textbf{a}]=\int{\cal D}[\Delta]e^{iS[\Delta,\textbf{a}]},
\end{equation}
with the effective action
\begin{equation}
\label{action}
iS[\Delta,\textbf{a}]
=
\Tr\ln\check{{\cal G}}^{-1}
-
\frac{2i\nu}{\lambda}\int d\br \, dt \left(\Delta^*_\text{cl}\Delta_\text{q}+\Delta^*_\text{q}\Delta_\text{cl}\right).
\end{equation}
Here $\Tr$ is the full operator trace, and $\check{{\cal G}}^{-1}$ is an operator in the Nambu and Keldysh spaces. Its structure in the Nambu space has the following form:
\begin{equation}
\check{{\cal G}}^{-1}
=
\begin{pmatrix}
i\partial_t-\check{H}_0 & \check\Delta\\
-\check\Delta^\ast & -i\partial_t-\check{H}^\ast_0
\end{pmatrix}_\text{N}.
\end{equation}
The operator $\check H_0$ is obtained from Eq.~\eqref{H0} by replacing $\textbf{a}$ by $\check{\textbf{a}}$, and the Keldysh matrices $\check{\Delta}$ and $\check{\textbf{a}}$ are given by
\be
\check\Delta
=
\Delta_\text{cl} \sigma_0 + \Delta_\text{q} \sigma_1
, \qquad
\check{\textbf{a}} 
=
\textbf{a}_\text{cl} \sigma_0 + \textbf{a}_\text{q} \sigma_1 ,
\ee
where $\sigma_i$ stand for the Pauli matrices in the Keldysh space.

In the normal state ($\Delta=0$), the Green function becomes diagonal in the Nambu space, with the particle ($G$) and hole ($\widetilde{G}$) components:
\begin{equation}
\label{G-diag-N}
{\cal G} =
\begin{pmatrix}
G & 0 \\
0 & \widetilde{G}
\end{pmatrix}_\text{N} .
\end{equation}
At equilibrium with the temperature $T$, the particle component is diagonal in the energy space with:
\begin{subequations}
\label{G-K-all}
\be
\label{G-K}
G_E =
\begin{pmatrix}
G^R_E & G^K_E \\
0 & G^A_E
\end{pmatrix}_\text{K} ,
\quad
G^K_E = F_E (G^R_E-G^A_E) ,
\ee
with $F_E=\tanh(E/2T)$ being the fermionic equilibrium distribution function. 
The hole component has the same structure in the Keldysh space and is related to the particle
component by the energy inversion:
\begin{equation}
\widetilde{G}^{R(A)}_E=-G^{A(R)}_{-E},
\quad 
\widetilde{G}^{K}_E=-G^{K}_{-E}.
\end{equation}
\end{subequations}

Gaussian fluctuations of the order parameter field governed by the action \eqref{action} are described by the fluctuation propagator 
\be
\label{L-def}
L_{ij}(\omega,q)=2i\nu\left<\Delta_i(\omega,q)\Delta_j^*(-\omega,-q)\right> .
\ee 
Its structure in the Keldysh space at equilibrium reads:
\be
L(\omega,q) =
\begin{pmatrix}
L^K_\omega & L^R_\omega \\
L^A_\omega & 0
\end{pmatrix}_\text{K}
,
\quad
L^K_\omega = B_\omega (L^R_\omega-L^A_\omega) ,
\ee
where $B(\omega)=\coth(\omega/2T)$ is the bosonic equilibrium distribution function, and $L^A_\omega=(L^R_\omega)^*$.

\begin{figure}
\includegraphics[width=\columnwidth]{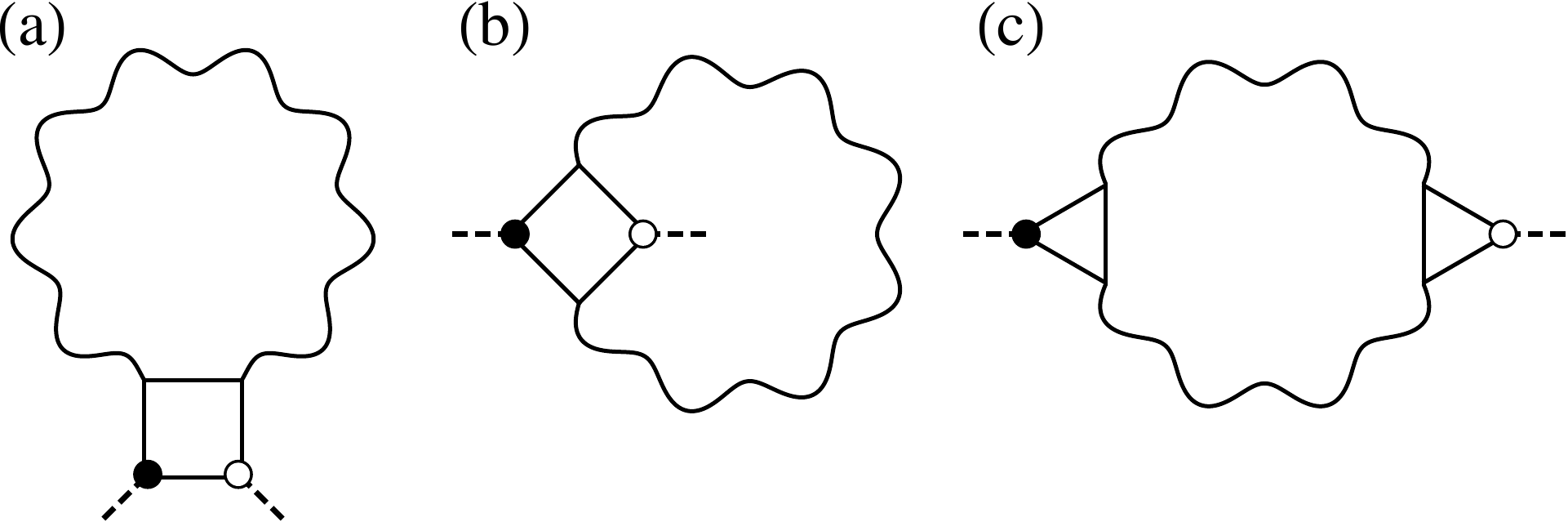}
\caption{Diagrams that determine the fluctuation correction to conductivity: (a) density of states (DOS), (b) Maki-Thompson (MT), (c) Aslamazov-Larkin (AL) contributions.
The wavy line is the fluctuation propagator, the solid line is the Green function.
Solid (open) dots indicate quantum (classical) current vertices.
The diagrams should be averaged over disorder.
}
\label{fig:Diagrams}
\end{figure}

\subsection{Conductivity correction}

In the Keldysh formalism, the linear response conductivity tensor is calculated by taking the second derivative with respect to the source fields $\textbf{a}$:
\begin{equation}
\delta\sigma_{\alpha\beta}(\omega)=\frac{Q_{\alpha\beta}(\omega)}{\omega}
,\qquad 
Q_{\alpha\beta}=-\frac{1}{2}\frac{\delta^2Z[\textbf{a}]}{\delta a^\alpha_\text{q} \delta a^\beta_\text{cl}}\bigg|_{\textbf{a}=0}.
\end{equation}

The fluctuation correction to conductivity in the one-loop (with respect to the fluctuation propagator) approximation is given by the three standard skeleton diagrams shown in Fig.~\ref{fig:Diagrams}: 
(a) DOS, (b) MT, and (c) AL \cite{LV-book}.
The kernel $Q(\omega)$ is given by the sum of the three corresponding contributions
\begin{equation}
\label{Q-def}
Q(\omega)=Q^\text{DOS}(\omega)+Q^\text{MT}(\omega)+Q^\text{AL}(\omega).
\end{equation}

The most complicated technical part of the calculation is the procedure of disorder averaging that is discussed in Appendices \ref{A:Initial expression} and \ref{A:Disorder averaging}. The general expression for $\delta\sigma$ will be presented in Sec.~\ref{SS:General}, and meanwhile we introduce the necessary ingredients.

\begin{figure}
\includegraphics[width=0.9\columnwidth]{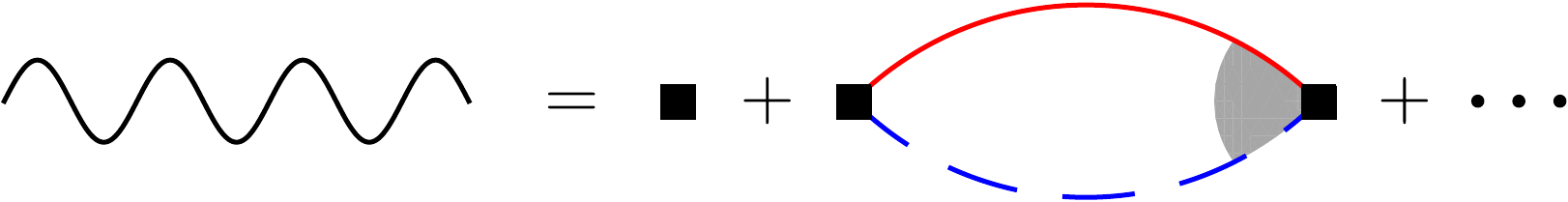}
\vskip 1mm
\caption{Equation for the disorder averaged fluctuation propagator. Solid (red) line is the retarded and dashed (blue) line is the advanced Green's function. The black dot denotes the electron-electron interaction vertex, and the gray sector stands for the disorder-dressed vertex shown in Fig.~\ref{fig:Vert}. 
}
\label{fig:Propagator}
\end{figure}

\begin{figure}[b]
\includegraphics[width=0.9\columnwidth]{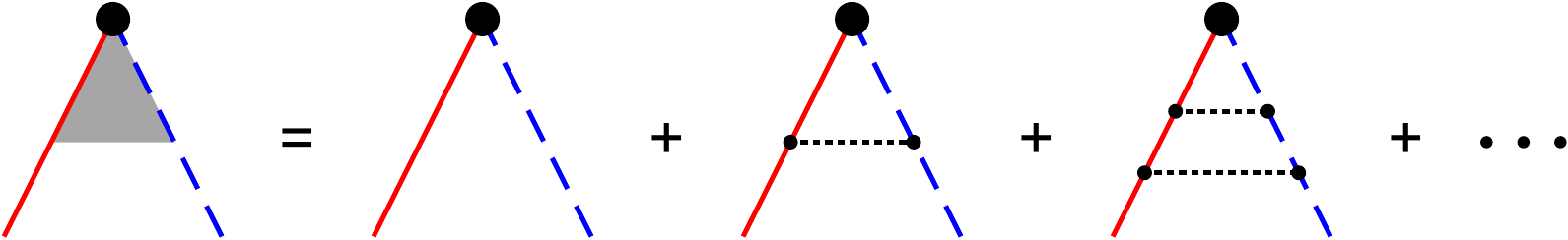}
\caption{Vertex correction $\gamma$ due to the impurity ladder. Black dotted lines denote the impurity correlator \eqref{<UU>}.
}
\label{fig:Vert}
\end{figure}

\subsection{Ingredients}
\label{SS:Ingredients}

The general expression for the fluctuation correction involves two basic ingredients: the fluctuation propagator $L^R(\omega,q)$ and the functions $f_m(\omega,q)$, $g_m(\omega,q)$ obtained from disorder averaging of the blocks of Green's functions in Fig.~\ref{fig:Diagrams}. We discuss these objects below.

\subsubsection{\textbf{Fluctuation propagator}}

Fluctuation propagator is defined as the sum of diagrams shown in Fig.~\ref{fig:Propagator}, which should also be averaged over disorder. In the leading approximation, the averaging should be performed in each bubble independently. The result is expressed in terms of the vertex correction
(see Fig.~\ref{fig:Vert}):
\be
\label{gamma-def}
\gamma^R(\omega,q) = \frac{1}{1-f^R_1(\omega,q)},
\ee
where $f_1(\omega,q)=\int G^R_{E+\omega}(p+q)G_E^A(p) (dp)/2\pi\nu\tau$ is the primitive step of the ladder.
In the limit $q\ll p_F$, it is given by
\be
\label{f1-def}
f_1^R(\omega,q)=\left<\frac{1}{1-i\omega\tau+iqln_x}\right>
_{\mathbf n} ,
\ee
where the averaging is taken over the unit $d$-dimensional vector $\mathbf{n}$, $l=v_F\tau$ is the mean-free path and $v_F$ is the Fermi velocity. 
Explicit expression for $f_1$ depends on the space dimensionalilty.
In the 2D and 3D cases, it is given by
\be
\label{f1-2D-3D}
f_1^R(\omega,q)
=
\begin{cases}
\displaystyle
\frac{1}{\sqrt{(1-i\omega\tau)^2+q^2l^2}} , & \text{2D} ; 
\\[12pt]
\displaystyle
\frac{\arctan[ql/(1-i\omega\tau)]}{ql} , & \text{3D} .
\end{cases}
\ee
In what follows we will work in the 2D case and use the corresponding expression for $f_1$.

The retarded fluctuation propagator $L^R(\Omega,q)$ can be written as the sum over Matsubara energies $\varepsilon_n=2\pi T(n+1/2)$ \cite{LV-book}:
\be
\label{LR-def}
\frac{1}{L^R(\Omega,q)}
=
\epsilon
+
4\pi T
\sum_{n=0}^\infty\left[
\frac{1}{2\varepsilon_n}
-
\frac{\tau f^R_1(2 i \varepsilon_n+\Omega,q)}{1-f^R_1(2 i \varepsilon_n+\Omega,q)} 
\right]
,
\ee
where 
\be
\epsilon = \ln(T/T_c) .
\ee
It's worth emphasizing that Eq.\ \eqref{LR-def} is consistent with the definition (\ref{L-def}) and does not contain the typical factor of $-1/\nu$ inherited from the interaction term in Eq.~(\ref{H}). We find it convenient to follow the present notation, which renders $L$ dimensionless and simplifies intermediate expressions.

The general form \eqref{LR-def} of the fluctuation propagator is quite cumbersome. It can be significantly simplified in the two partially overlapping limits: (i) in the diffusive regime at arbitrary temperatures, and (ii) near the transition at arbitrary disorder strength.

\begin{itemize}

\item
\emph{In the diffusive limit}\ ($ql\ll 1$ and $\Omega\tau\ll 1$), 
the function $f_1$ is close to unity, 
$f_1^R(\omega,q)\approx1-i\omega\tau+Dq^2\tau$, and the fluctuation propagator takes a simple form:
\be
\label{Diff-prop-def}
\frac{1}{L^R(\Omega,q)}
=\epsilon+\psi\left(\frac12+\frac{Dq^2-i\Omega}{4\pi T}\right)
-\psi\left(\frac12\right) ,
\ee
where $\psi(z)$ is the digamma function.

\item
\emph{Near the transition}\ ($T\to T_c$), the main contribution to conductivity comes from small momenta and frequencies. The fluctuation propagator then takes the so-called local form regardless of the value of $T\tau$ \cite{LV-book}:
\be
\label{LR-cross-near}
\frac{1}{L^R(\Omega,q)}
=
\epsilon+\xi^2(T\tau) q^2-\frac{i\pi}{8T}\Omega ,
\ee
where $\xi(T\tau)$ is the disorder-dependent 
coherence length. It interpolates between the dirty, $\xi_d^2=\pi D/8T$, and clean, $\xi_c^2=7\zeta(3)D/16\pi^2\tau T^2$, limits according to the general expression \cite{Gorkov,AL}:
\be
\label{xi-def}
\xi^2(T\tau)= D\tau \, {\cal{F}}(1/2), 
\ee
where the function ${\cal F}$ is defined as
\be
{\cal{F}}(z)
=
\psi(z) 
+ \frac{\psi'(z)}{4\pi T\tau}
- \psi\left(z+\frac{1}{4\pi T\tau}\right) .
\label{F-def}
\ee
Note that $\xi(T\tau)$ remains finite at $T\to T_c$.
It is the correlation length $\xi(T\tau)/\sqrt{\epsilon}$ [see Eq.\ \eqref{LR-cross-near}] that becomes singular near the transition.

\end{itemize}

\color{black}

\subsubsection{\textbf{Disorder averaged blocks}}

The procedure of disorder averaging of the blocks of electron Green's functions is outlined in Appendix \ref{A:Disorder averaging}. In calculating the integrals over the Fermi sphere, one encounters the following expressions:
\begin{subequations}
\label{fg}
\begin{gather}
\label{f-def}
f^R_m(\omega,q)
=
\left< \frac{1}{(1-i\omega\tau+iqln_x)^m} \right>_\mathbf{n} ,
\\
\label{g-def}
g^R_m(\omega,q)
=
\left< \frac{in_x}{(1-i\omega\tau+iqln_x)^m} \right>_\mathbf{n} .
\end{gather}
\end{subequations}
The functions $f_m(\omega,q)$ are generalizations of the function $f_1(\omega,q)$ in Eq.~\eqref{f1-def}. They appear after disorder averaging of the blocks for the DOS and MT diagrams. The functions $g_m(\omega,q)$ containing an additional $n_x$ in the numerator originate from averaging of the vector triangular vertex for the AL diagram.
As usual, $f_m^A(\omega,q)=[f_m^R(\omega,q)]^*$ and $g_m^A(\omega,q)=[g_m^R(\omega,q)]^*$.

\subsection{DC conductivity at arbitrary disorder}
\label{SS:General}

In the zero-frequency limit, the general expression for the conductivity correction can be conveniently represented as a sum of the contributions with one and two fluctuation propagators:
\be
\label{ds=ds+ds}
\delta\sigma
=
\delta\sigma^{(1)} + \delta\sigma^{(2)} .
\ee
The term with one proparator, $\delta\sigma^{(1)}$, accounts for the DOS and MT diagrams and also contains a part of the AL diagram (AL1) naturally transformed to a single $L$ via integration by parts, as described in Appendix \ref{AA:AL}.
The remaining part of the AL correction (AL2) yields a two-propagator contribution $\delta\sigma^{(2)}$.

\begin{widetext}
After some algebra, the expression for $\delta\sigma^{(1)}$ can be conveniently represented as 
\be
\label{ds-via-Sigma}
\delta\sigma^{(1)}
=
2 \pi D \tau ^2 
\int \frac{d\Omega}{2\pi}
\int \frac{d^dq}{(2\pi)^d}
\Bigl\{
B_\Omega
\Im \left[ L^R(\Omega,q) \Sigma^R(\Omega,q) \right]
+ B'_\Omega
\Im \left[ L^R(\Omega,q) \right] \Sigma^Z(\Omega,q)
\Bigr\} ,
\ee
where $\Sigma^{R,Z}(\Omega,q)$ denote blocks of electron Green's functions averaged over disorder.
The square block of the DOS diagram (a) in Fig.~\ref{fig:Diagrams} yields $\Sigma_\text{DOS}$ (without an additional cooperon; in the diffusive limit equivalent to the diagrams 5--8 from Ref.~\cite{LV-book}) and $\Sigma_\text{DOS(C)}$ (with an additional cooperon; in the diffusive limit equivalent to the diagrams 9--10 from Ref.~\cite{LV-book}).
The square block of the MT diagram (b) yields the regular contribution $\Sigma_\text{RMT}$ [without an additional cooperon, see diagrams (a)--(e), (g), (h) in Fig.~\ref{fig:Blocks-MT}; in the diffusive limit it is contained in the diagram 2 from Ref.~\cite{LV-book}], the anomalous contribution $\Sigma_\text{AMT}$ [the diagram (f) in Fig.~\ref{fig:Blocks-MT}; in the diffusive limit it is  contained in the diagram 2 from Ref.~\cite{LV-book}],
and $\Sigma_\text{MT(C)}$ [with an additional cooperon, see diagrams (b')--(h') in Fig.~\ref{fig:Blocks-MT}; in the diffusive limit it is equivalent to the diagrams 3, 4 from Ref.~\cite{LV-book}].

The functions $\Sigma^{R,Z}(\Omega,q)$ are given explicitly by the following expressions:
\begin{subequations}
\label{SZR}
\begin{gather}
\label{S_R-DOS+RMT}
\Sigma^R_\text{DOS+RMT}(\Omega,q)
=
\int \frac{dE}{2\pi}
F'_E 
\left[
f_1^A - f_1^A (\gamma^A)^2 - 2 f_2^A \gamma^A - 3 f_3^A (\gamma^A)^2
\right] ,
\\
\label{S_Z-DOS+RMT}
\Sigma^Z_\text{DOS+RMT}(\Omega,q)
=
2 \Re
\int \frac{dE}{2\pi}
F_E 
\left[
f_1^A + f_1^A (\gamma^A)^2
- 2 f_2^A \gamma^A + f_3^A (\gamma^A)^2
\right] ,
\\
\label{S_R-MT(C)}
\Sigma^R_\text{MT(C)}(\Omega,q)
=
2 
\int \frac{dE}{2\pi}
F'_E
\left[
f_1^A g_1^A g_2^A (\gamma^{A})^2
+ f_1^A (g_2^A)^2 (\gamma^{A})^3
\right] ,
\\
\label{S_Z-MT(C)}
\Sigma^Z_\text{MT(C)}(\Omega,q)
=
4 \Re
\int \frac{dE}{2\pi}
F_E
f_1^A g_1^A g_2^A (\gamma^{A})^2 ,
\\
\label{S_R-DOS(C)}
\Sigma^R_\text{DOS(C)}(\Omega,q)
=
\int \frac{dE}{2\pi}
F'_E
\left[
f_1^A (g_2^{A})^2 (\gamma^A)^3
- f_1^A (g_1^{A})^2 \gamma ^A 
\right] ,
\\
\label{S_Z-DOS(C)}
\Sigma^Z_\text{DOS(C)}(\Omega,q)
=
- 2 \Re
\int \frac{dE}{2\pi}
F_E
\left[ 
f_1^A (g_2^A)^2 (\gamma^A)^3 
+
f_1^A (g_1^A)^2 \gamma^A
\right] ,
\\
\label{S_Z-AMT}
\Sigma^Z_\text{AMT}(\Omega,q)
=
-4 \Re
\int \frac{dE}{2\pi}
F_E f_1 ^A \gamma^R \gamma^A ,
\\
\label{S_R-AL1}
\Sigma^R_\text{AL1}(\Omega,q)
=
\frac{2}{l}
\frac{1}{q^{d-1}} \partial_q 
\bigg\{
q^{d-1}
\int \frac{dE}{2\pi}
F'_E
\left[ g_2^{A} (\gamma^{A})^2 + 2 g_1^{A} \gamma^{A} \right] \bigg\}
,
\end{gather}
\end{subequations}
where the arguments of functions $f_m$, $g_m$ and $\gamma$ are $f_m(2E-\Omega,q)$, etc.,
and $\partial_q = \partial/\partial|q|$ is the derivative with respect to the absolute value of the momentum.
The blocks absent in the list (\ref{SZR}) are zeros:
$\Sigma^R_\text{AMT} = \Sigma^Z_\text{AL1} = 0$.

To regularize the infra-red divergency of the AMT correction in the dimensionalities $d\leq2$, one should take into account a finite dephasing time $\tau_\phi$ \cite{MT1,MT2}. 
We will neglect dephasing due to scattering on superconducting fluctuations \cite{Patton,KellerKorenman} which is characterized by an energy-dependent dephasing rate \cite{Reizer92} and assume, e.g., the standard Coulomb-mediated dephasing \cite{AAK}.

Remarkably, the two-propagator contribution, $\delta\sigma^{(2)} = \delta\sigma^\text{AL2}$, can be expressed solely in terms of two $L$'s and their derivatives [which generate the blocks, according to Eq.~(\ref{L^2-L})]:
\be
\label{sigma-AL2-v2}
\delta\sigma^{(2)}
=
\frac{1}{2d} 
\int \frac{d\Omega}{2\pi}
\int \frac{d^dq}{(2\pi)^d}
B'(\Omega)
\Bigl[
2
\left( 
L^R \partial_q [L^R]^{-1} - L^A \partial_q [L^A]^{-1}
\right)^2
-
L^R L^A
\left( 
\partial_q [L^R]^{-1}
-
\partial_q [L^A]^{-1}
\right)^2
\Bigr] .
\ee

Equations (\ref{ds=ds+ds})--(\ref{sigma-AL2-v2}) describe dc fluctuation conductivity 
in units of $e^2/\hbar$ 
for arbitrary temperatures, $T>T_c$, disorder 
scattering times $\tau$,
and space dimensionality $d$. We analyze them below 
in the 2D geometry.

\end{widetext}

\section{DC conductivity in the diffusive regime}
\label{S:Diffusive-regime}

In this Section we analyze the fluctuation correction 
at arbitrary temperatures in the diffusive regime, $T\tau\ll1$ [regions (a) and (b) in Fig.\ \ref{fig:phasediagram}].
In terms of length scales that corresponds to the inequality $l\ll \xi_c$, when the coherence length is given by the dirty-limit expression $\xi_d=\sqrt{\pi D/8T} \sim \sqrt{\xi_cl}$, see Eq.~\eqref{xi-def}. In the diffusive regime, $f_m\approx 1$, $g_m\approx mql/d$, and the generalized cooperon acquires the standard diffusive form:
\be
\label{gamma-dif}
\gamma^{R(A)}(\omega,q) = \frac{1}{\tau(Dq^2\mp i\omega)}.
\ee
Then the integrals over $E$ in Eqs.\ (\ref{SZR}) except for the $\Sigma^Z_\text{AMT}$ can be calculated by the residues of $F_E$, since the integration contour can be deformed to avoid singularities of $f$'s and $\gamma$'s. The resulting expression is then expressed in terms of the digamma function. On the other hand, the AMT contribution (\ref{S_Z-AMT}) contains a part determined by the poles of $\gamma^R\gamma^A$.

\subsection{General expression in 2D}
\label{S:dif-gen}

In the two-dimensional case, the fluctuation correction to the conductivity can be written as
\be
\label{sigma-varsigma}
\delta\sigma
=
\frac{1}{4\pi^2}
\int_0^\infty dx
\int_{-\infty}^\infty dy \,
\varsigma(x,y) ,
\ee
where the dimensionless variables $x$ and $y$ are related to the
momentum and frequency of the fluctuation propagator as
\be
\label{xy-dif}
x = \frac{Dq^2}{4\pi T} ,
\qquad
y = \frac{\Omega}{4\pi T} .
\ee

Following Ref.~\cite{Petkovic}, we find it convenient to express the result in terms of the function $G(z)$ of the complex variable $z=x+iy$:
\be
\label{G(z)}
G(z) 
=
\frac{1}{L^A(\Omega,q)}
= 
\epsilon 
+
\psi\left(\frac12+z\right)
- 
\psi\left(\frac12\right)
,
\ee
which determines both the fluctuation propagator, 
and the blocks $\Sigma$ in Eqs.~\eqref{SZR}.
The contribution of the diagrams to $\varsigma(x,y)$ are listed below:
\begin{gather}
\varsigma^\text{DOS+RMT}
=
- 4 b \Im \frac{G''}{G}
- 2 b' \frac{\Im G \Im G'}{|G|^2} ,
\\
\varsigma^\text{DOS(C)+MT(C)}
=
- 3
b \, x \Im \frac{G'''}{G}
-
b' x \,
\frac{\Im G\Im G''}{|G|^2} ,
\\
\varsigma^\text{AL1}
=
4 b \Im
\frac{G'' + x G'''}{G} ,
\\
\label{varsigma-AMT}
\varsigma^\text{AMT}
=
- 
\frac{2b'}{x+x_*}
\frac{\Im^2 G}{|G|^2} ,
\\
\label{varsigma-AL2}
\varsigma^\text{AL2}
=
2 b' x
\biggl(
\frac{\Im\nolimits^2G'}{|G|^2}
-
2 \Im\nolimits^2\frac{G'}{G}
\biggr) ,
\end{gather}
where $G=G(z)$, $b=b(y)=\coth2\pi y$, $x_*=(4\pi T\tau_\phi)^{-1}$ is the dimensionless dephasing rate, and primes indicate derivatives with respect to the corresponding argument.

We see that the mathematical structure of the expressions for the DOS, RMT, DOS(C), MT(C), and AL1 contributions is similar. Therefore we find it natural to combine them into a single quantity
\be
\varsigma^{\text{REG}}
=
\varsigma^{\text{DOS+RMT}}+\varsigma^{\text{DOS(C)+MT(C)}}+\varsigma^{\text{AL1}}
,
\ee
which after partial cancellations acquires a fairly simple form:
\be
\label{varsigma-dif-reg}
\varsigma^\text{REG}
=
b \, x \Im \frac{G'''}{G} 
- b' \frac{\Im G \Im (2 G' + xG'')}{|G|^2} .
\ee
As a result, the total correction can be written in the form (\ref{sigma-varsigma}) with 
\be
\label{varsigma-3}
\varsigma
=
\varsigma^\text{REG}
+
\varsigma^\text{AL2}
+
\varsigma^\text{AMT}
,
\ee
where $\varsigma^\text{REG}$, $\varsigma^\text{AL2}$, and $\varsigma^\text{AMT}$ are given by Eqs.~(\ref{varsigma-dif-reg}), (\ref{varsigma-AL2}) and (\ref{varsigma-AMT}), respectively.

\subsection{Comparison with previous results}

Our expression for the fluctuation correction in the 2D diffusive case 
[given by Eqs.~(\ref{sigma-varsigma}), (\ref{varsigma-AMT}), (\ref{varsigma-AL2}), (\ref{varsigma-dif-reg}) and (\ref{varsigma-3})] 
exactly coincides with the zero-field result obtained by Tikhonov \emph{et al.} in a different approach \cite{Tikhonov}. In Ref.~\cite{Tikhonov}, the total correction is represented as a sum of three terms, $\delta\sigma^\text{(dos)}+\delta\sigma^\text{(sc)}+\delta\sigma^\text{(an)}$.
Direct comparison shows that their $\delta\sigma^\text{(an)}$ is equal to our $\delta\sigma^\text{AMT}$. 
The other two terms, $\delta\sigma^\text{(dos)}$ and $\delta\sigma^\text{(sc)}$, 
are given by Eqs.~(\ref{Tikhonov-def-dos}) and (\ref{Tikhonov-def-sc}), respectively.
Though these terms are not separately equal to $\delta\sigma^\text{REG}$ and $\delta\sigma^\text{AL2}$,
we demonstrate in Appendix \ref{A:Comp2Kostya} that their overall contribution is same: 
$\delta\sigma^\text{(dos)}+\delta\sigma^\text{(sc)}
=
\delta\sigma^\text{REG}+\delta\sigma^\text{AL2}$.
Thus we have established a complete equivalence between the standard diagrammatic approach (in the Keldysh form) and the method based on the 
Usadel equation 
in the field of a fluctuating order parameter (also in the Keldysh form) developed in Ref.~\cite{Tikhonov}.

In the language of the Keldysh sigma model, the fluctuation conductivity was studied by Petkovi\'c and Vinokur~\cite{Petkovic}, where the total correction was written in a different way: $\delta\sigma_\text{DOS}+\delta\sigma_\text{AL}+\delta\sigma_\text{MT}$. 
Their MT correction, $\delta\sigma_\text{MT}$, coincides with $\delta\sigma^\text{(an)}$ of Ref.~\cite{Tikhonov} and with our $\delta\sigma^\text{AMT}$.
However the expressions for $\delta\sigma_\text{DOS}+\delta\sigma_\text{AL}$ in Ref.~\cite{Petkovic} are too cumbersome, and we did not compare them with our $\delta\sigma^\text{REG}+\delta\sigma^\text{AL2}$. Nevertheless, our asymptotic expansion of the correction in the region $\gamma\ll\epsilon\ll1$ [see Eq.~(\ref{sigma-tot-diff-near}) below] fully agrees with their result, which is a strong argument in favor of a full equivalence between our theory and the theory of Ref.~\cite{Petkovic}.

Finally, we mention the work by Glatz \emph{et al.}\ \cite{Glatz2011}, where the fluctuation correction at arbitrary temperatures and magnetic fields was calculated within the standard Matsubara diagrammatic technique. As we fully coincide with Ref.~\cite{Tikhonov} in the zero-field regime, we do not reproduce the results of Ref.~\cite{Glatz2011} (results of Refs.~\cite{Glatz2011,Tikhonov} are known to be inconsistent).

\subsection{Vicinity of the transition, $T\to T_c$}
\label{SS:diff-close}

In the vicinity of $T_c$ [region (a) in Fig.\ \ref{fig:phasediagram}], the general expressions for $\delta\sigma^\text{REG}$, $\delta\sigma^\text{AL2}$, and $\delta\sigma^\text{AMT}$ can be simplified. 
While the leading AL and AMT terms are well established \cite{LV-book}, the behavior of the subleading terms as a function of $\epsilon$ and 
$\gamma=(\pi/8)(T\tau_\phi)^{-1}$ 
is still a subject of controversy \cite{Tikhonov,Petkovic}.
Keeping the terms which are singular in the limit $\epsilon, \gamma \to 0$, we get with the accuracy of ${\cal O}(1)$:
\begin{gather}
\label{AL2}
\delta\sigma^\text{AL2}
=
\frac{1}{16\epsilon} 
+
\dots
,
\\
\label{sigma-reg-diff-close}
\delta\sigma^\text{REG}
=
- \frac{7 \zeta(3)}{\pi^4} \ln \frac{1}{\epsilon} +
\dots
,
\\
\label{sigma-AMT-res-diff-close}
\delta\sigma^\text{AMT}
=
\frac{1}{8} \frac{\ln(\epsilon/\gamma)}{\epsilon-\gamma} 
+
h\left(\frac{\gamma}{\epsilon}\right)
- 
\left[ \frac{7\zeta(3)}{\pi^4} + \kappa \right]
\ln\frac{1}{\gamma}
+
\dots
,
\end{gather}
where the numerical constant $\kappa=0.1120$ and the function $h(x)$ is given by
\be
h(x) 
= 
\frac{7\zeta(3)}{2 \pi ^4}
\frac{\ln(1/x) - 1 + x}{(1-x)^2} .
\ee

The coefficient in front of $\ln(1/\gamma)$ in the AMT correction (\ref{sigma-AMT-res-diff-close}) contains two terms of different origin, as discussed in Appendix~\ref{A:AMT}. The one proportional to $\zeta(3)$ comes from small $y$ [see Eq.\ \eqref{xy-dif}] corresponding to the frequency $\Omega\sim T-T_c$, where the function $G(z)$ in Eq.~(\ref{varsigma-AMT}) can be expanded in a power series. On the contrary, the term with $\kappa$ is determined by the integral (\ref{kappa}), where $y\sim1$ (corresponding to $\Omega\sim T_c$) are important and the digamma function should be left unexpanded. The relevance of this parameter region was pointed out in Ref.~\cite{Petkovic}, where the parameter $\kappa$ (equal to $7\zeta(3)(c-1)/\pi^4$ in the notations of Ref.~\cite{Petkovic}) was estimated numerically with the 4\% accuracy.

In the relevant case of weak phase breaking, $\gamma\ll\epsilon\ll1$, we find for the total correction:
\be
\label{sigma-tot-diff-near}
\delta\sigma
=
\frac{1+2\ln(\epsilon/\gamma)}{16\epsilon}
-
\frac{21\zeta(3)}{2\pi ^4} \ln\frac{1}{\epsilon}
- 
\left[ \frac{7\zeta(3)}{2\pi^4} + \kappa \right]
\ln\frac{1}{\gamma}
+
\dots
\ee
Here the first term is the sum of the standard AL and AMT corrections in the leading approximations, and the coefficient $c=-21/2$ in front of the subleading term $\zeta(3)\ln(1/\epsilon)/\pi^4$ is a sum of 
\be
\label{ccc}
c^\text{REG}=-7, 
\quad
c^\text{AL2}=0, 
\quad
c^\text{AMT}=-7/2 .
\ee
Equation~(\ref{sigma-tot-diff-near}) exactly coincides with the expression derived in Ref.~\cite{Petkovic}, where the coefficient $c=-21/2$ is obtained as a sum of $c_\text{DOS}=-21$, $c_\text{AL}=14$, and $c_\text{MT}=-7/2$.
The coefficients $c^\text{(dos)}=-7$ and $c^\text{(sc)}=0$ found in Ref.~\cite{Tikhonov} coincide with our $c^\text{REG}=-7$ and $c^\text{AL2}=0$, respectively, whereas their $c^\text{(an)}$ contains an error corrected in Ref.~\cite{Petkovic}.

In this Section we present our results in terms of $\delta\sigma^\text{REG}$, $\delta\sigma^\text{AL2}$, and $\delta\sigma^\text{AMT}$, due to their compact analytical form. One can easily show that the initial diagrams produce the following contributions to the coefficients in front of the logarithmic term:
\be
c^\text{DOS}=-14, 
\quad
c^\text{AL}=14, 
\quad
c^\text{MT}=-7-7/2 
\ee
(where $c^\text{RMT}=-7$ and $c^\text{AMT}=-7/2$). 
In the standard diagrammatic calculation in the Matsubara technique~\cite{LV-book}, the contribution of the MT diagram is often represented as a sum of the regular and anomalous terms, $\delta\sigma^\text{MT,reg}$ and $\delta\sigma^\text{MT,an}$. Such decomposition is purely technical, naturally arising in the process of analytic continuation. It is similar but yet different from our splitting of the MT diagram into the RMT and AMT terms, which is also a matter of technical convenience. In the vicinity of the transition, $\delta\sigma^\text{MT,reg}$ grows logarithmically with $c^\text{MT,reg}=-14$, leading to the doubling of the DOS correction \cite{LV-book}. Note however that an accurate extraction of the subleading logarithmic term from the anomalous part of the MT contribution should give $c^\text{MT,an}=7/2$, restoring the correct overall coefficient $c^\text{MT}=-21/2$.

\subsection{Far above the transition, $T\gg T_c$}
\label{SS:diff-large-T}

At high temperatures, $\epsilon=\ln (T/T_c) \gg 1$ [region (b) in Fig.\ \ref{fig:phasediagram}], the main temperature-dependent contribution to the conductivity is given by the term $\delta\sigma^{\text{REG}}$ coming from large momenta and frequencies ($x,y\sim e^\epsilon\gg 1$). Formally, the correction diverges at large momenta, $x\gg e^\epsilon$, and is usually cut at the upper limit of the diffusion region $x\sim1/(T\tau)$, leading to the known result \cite{Altshuler-Varlamov}:
\begin{equation}
\label{sigma-reg-diff-far}
\delta\sigma^{\text{REG}} 
=
-
\frac{1}{2\pi^2}
\ln\frac{\ln 1/T_c\tau}{\epsilon}
+ {\cal O}\left(\epsilon^{-1}\right) .
\end{equation}
In the ballistic region, as we will see in Sec.~\ref{SS:ball-gen}, the ultra-violet divergency becomes even stronger, leading to a large but temperature-independent and therefore unmeasurable contribution.

The functions $\varsigma^{\text{AMT}}$ and $\varsigma^{\text{AL2}}$ are proportional to $b'$, therefore the corresponding conductivity corrections come from $x\sim 1$ and $y\sim 1$. Hence they are additionally suppressed in the parameter $1/\epsilon$ compared to $\delta\sigma^{\text{REG}}$:
\begin{gather}
\label{sigma-MT-diff-far}
\delta\sigma^{\text{AMT}}
=
\frac{1}{12}
\frac{\ln 1/\gamma}{\epsilon^2}
+ {\cal O}\left(\epsilon^{-3}\right) ,
\\
\delta\sigma^{\text{AL2}}
=
\frac{\varkappa}{\epsilon^2}
+ {\cal O}\left(\epsilon^{-3}\right) ,
\end{gather}
where a large factor $\ln 1/\gamma$ in the AMT correction is due to the singularity at small momenta, and a numerical constant $\varkappa=0.0150$ is determined by the integral
\begin{equation}
\varkappa
=
\int_0^\infty dx\int_{-\infty}^{\infty}dy\frac{x \Im^2\psi'(1/2+x+i y)}{\pi \sinh^2 2\pi y} .
\end{equation}

For the sake of comparison with previous diagrammatic studies, it is instructive to look at the full contribution of the AL diagram, $\delta\sigma^{\text{AL}}=\delta\sigma^{\text{AL1}}+\delta\sigma^{\text{AL2}}$. In our representation, the term $\delta\sigma^{\text{AL1}}$ relegated to $\delta \sigma^{\text{REG}}$ appears to be much larger than $\delta\sigma^{\text{AL2}}$ and thus determines the leading asymptotics of the AL correction:
\begin{equation}
\label{sigma-AL-diff-far}
\delta\sigma^{\text{AL}}
=
\frac{1}{\pi^2}
\left(\frac{1}{\epsilon}-\frac{1}{\ln 1/T_c\tau}\right)
+ {\cal O}\left(\epsilon^{-2}\right) .
\end{equation}

Fluctuation conductivity at $T>T_c$ was first considered in Ref.~\cite{Altshuler-Varlamov}. We reproduce their results for the leading correction (\ref{sigma-reg-diff-far}) and the AL contribution (\ref{sigma-AL-diff-far}). At the same time, we cannot reproduce the partial contributions of individual diagrams in Ref.~\cite{Altshuler-Varlamov}. The discrepancy appears, e.~g., in the relative contributions of the DOS and MT diagrams crossed by additional cooperons:
Instead of their relation $2\delta\sigma^{\text{DOS(C)}}=-\delta\sigma^{\text{MT(C)}}$ we obtain
$2\delta\sigma^{\text{DOS(C)}}=\delta\sigma^{\text{MT(C)}}$.
Also, our AMT correction (\ref{sigma-MT-diff-far}) is $\pi^2/2$ times smaller than the one obtained in Ref.~\cite{Altshuler-Varlamov}.
Finally, our asymptotic expressions fully agree with the result of Ref.~\cite{Tikhonov}.

\section{DC conductivity in the ballistic regime}
\label{S:Ballistic-regime}

\subsection{General discussion}
\label{SS:ball-gen}

Consider now the fluctuation correction in the ballistic regime, $T\tau\gg1$, corresponding to the inequality $l\gg\xi_c$ [regions (c) and (d) in Fig.\ \ref{fig:phasediagram}]. In this case it is convenient to use the dimensionless momentum, $x$, and frequency, $y$, introduced according to
[cf.\ Eq.~(\ref{xy-dif})]
\be
\label{xy-ball}
x = \frac{q v}{4 \pi T}, 
\qquad
y = \frac{\Omega}{4 \pi T} .
\ee
In the two-dimensional case considered hereafter, the general expression (\ref{ds-via-Sigma}) for the one-propagator contribution can be written as 
\begin{multline}
\label{ds-via-Sigma-ball}
\delta\sigma^{(1)}
=
16\pi^2 T^3\tau^3
\int_0^\infty x \, dx
\int_{-\infty}^\infty dy
\\{}
\times
\left\{
b(y) 
\Im (L^R \Sigma^R)
+ \frac{b'(y)}{4\pi T}
\Im (L^R) \Sigma^Z
\right\} ,
\end{multline}
where $b(y)=\coth2\pi y$, as in Sec.\ \ref{S:dif-gen}.

Let us analyze the behavior of $\delta\sigma^{(1)}$ in the limit $\tau\to\infty$.
The blocks $f_n$ and $g_n$ defined in Eqs.~(\ref{fg}) scale as ${\cal O}(1/\tau^n)$. 
Then the most singular contributions among Eqs.~(\ref{SZR}) are given by $\Sigma^Z_\text{DOS+RMT}$ and $\Sigma^Z_\text{AMT}$, where one should replace $\gamma\to1$ and keep only $f_1$. Both terms behave as ${\cal O}(1/\tau)$ but are opposite in sign and exactly cancel each other. Retaining the next-to-leading terms ${\cal O}(1/\tau^2)$, we obtain
\begin{gather}
\label{SR-clean}
\Sigma^R
=
2 \int \frac{dE}{2\pi}
F'_E 
\left[
f_2^A - (f_1^A)^2
\right] 
+ {\cal O}(1/\tau^3) ,
\\
\label{SZ-clean}
\Sigma^Z
=
-4 \Re
\int \frac{dE}{2\pi}
F_E 
\left[
f_2^A + f_1^Rf_1^A 
\right] 
+ {\cal O}(1/\tau^3) ,
\end{gather}
where we have employed the identity 
$ \frac{1}{l}
\frac{1}{q^{d-1}} \frac{\partial}{\partial q} q^{d-1} g_1
=
f_2
$ 
(valid for any dimensinality $d$)
in transforming $\Sigma^R_\text{AL1}$.

The two-propagator AL2 contribution (\ref{sigma-AL2-v2}) and omitted terms of the MT contributions contain a smaller power of the large parameter $T\tau$ but have a stronger divergency at $T\to T_c$, see below.

The analytic structure of the integrands in $\Sigma^R$ and $\Sigma^Z$ [Eqs.\ \eqref{SZR}] is determined by (i) the poles of the distribution function $F_E$ and (ii) the singularities of $f_1^A\gamma^R\gamma^A$ in the AMT term $\Sigma^Z_\text{AMT}$. 
Indeed, the singularities due to $f_1^{R,A}$ and $\gamma^{R,A}$ in all the terms except for the AMT term are located either above of below the real axis and therefore can be avoided by a proper deformation of the integration contour. The corresponding integral over $E$ is then determined by the poles of $F_E$.

\begin{figure}
\includegraphics[width=0.9\columnwidth]{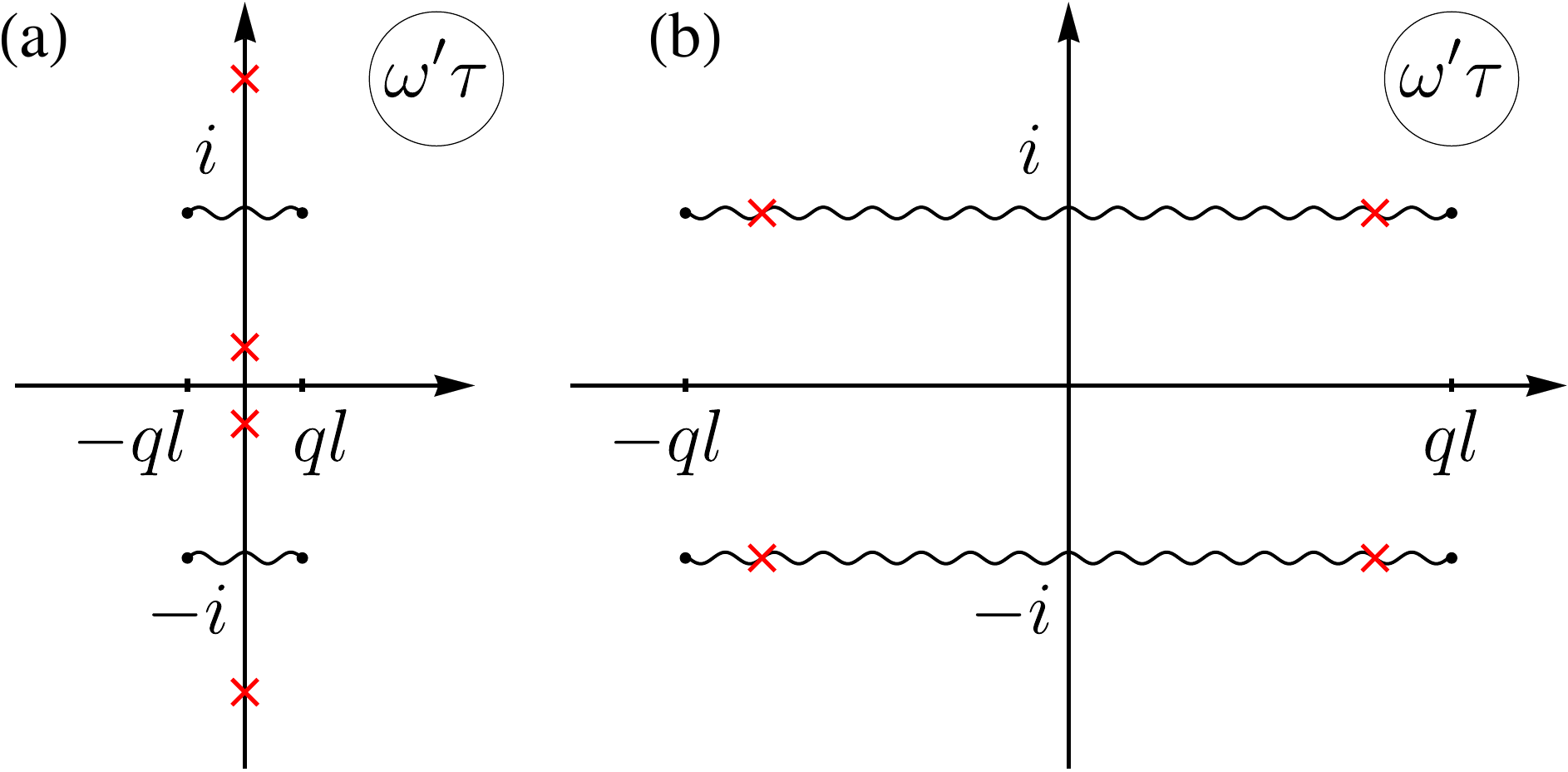}
\vskip 1mm
\caption{Singularities of the term $f_1^A(\omega',q)\gamma^R(\omega',q)\gamma^A(\omega',q)$ in the integrand of Eq.~(\ref{S_Z-AMT}) in the plane of complex 
frequency $\omega'=2E-\omega$: 
branch cuts of $f_1^{R,A}$ shown by wavy lines and poles of $\gamma^{R,A}$ shown by crosses. Panels (a) and (b) correspond to diffusive ($ql\ll 1$) and ballistic ($ql\gg 1$) momenta, respectively.}
\label{F:cuts}
\end{figure}

On the contrary, the AMT term (\ref{S_Z-AMT}) contains the product $f_1^A\gamma^R\gamma^A$ which has singularities both above and below the real axis: the branch cuts of $f^{R,A}_1$ and the poles of $\gamma^{R,A}$, see Fig.\ \ref{F:cuts}. The poles of cooperons $\gamma^{R,A}$ becomes essential only for small momenta, $ql\ll1$. The contribution of this region is smaller than the contribution of the region $ql\gg1$ [the term $f_1^Rf_1^A$ in Eq.\ \eqref{SZ-clean}] by the factor of $1/T\tau$, but is more singular at $\epsilon\to 0$. 
It will be analyzed in Sec.\ \ref{SS:clean-close}.

Since in the ballistic case the poles of $F_E$ are located much further from the real axis than other singularities in $\Sigma^Z_\text{AMT}$, it is natural to separate the AMT term into the contribution of the poles of the distribution function, $\Sigma^{Z, \text{AMT}}_\text{tanh}$, and the singularities of $f_1^A\gamma^R\gamma^A$, $\Sigma^Z_\text{sing}$. Hence we have three contributions: $\Sigma^R$, $\Sigma^Z_\text{tanh}$ (which includes $\Sigma^{Z, \text{AMT}}_\text{tanh}$ and all other $\Sigma^Z$) and $\Sigma^Z_\text{sing}$.
The functions $\Sigma^R$ [Eq.~(\ref{SR-clean-res})], $\Sigma^Z_\text{tanh}$ [Eq.~(\ref{SZ-clean-poles-res})] and $\Sigma^Z_\text{sign}$ [Eq.~(\ref{SZ-clean-sing-res})] are evaluated in Appendix \ref{A:Sigma-bal} in the leading order in $\tau$. 
Since both $\Sigma^R$ and $\Sigma^Z_\text{tanh}$ scale as $1/\tau^2$, the resulting $\delta\sigma$ is proportional to $\tau$, analogously to the bare Drude conductivity. 
Below we analyze the fluctuation correction near the transition and at high temperatures.

\subsection{Vicinity of the transition, $T\to T_c$}
\label{SS:clean-close}

The fluctuation propagator in the ballistic regime is given by Eq.~\eqref{LR-inv-ball}.
In the limit $T\to T_c$ [region (c) in Fig.\ \ref{fig:phasediagram}], the singular part of the fluctuation correction is determined by small momenta and frequencies that allows to expand assuming $x,y\ll1$:
\be
\label{LA-ball-near}
\frac{1}{L^A(\Omega,q)}
=
\epsilon+\frac{7\zeta(3)}{2}x^2+i\frac{\pi^2}{2}y ,
\ee
which is the ballistic limit of the general expression \eqref{LR-cross-near}.

We start with analyzing the leading contributions in $T\tau\gg1$.
According to Eq.~\eqref{LA-ball-near}, the singular part of $\delta\sigma$ comes from by $x\sim\sqrt\epsilon$ and $y\sim\epsilon$, corresponding to $q\sim\sqrt{\epsilon}\,T/v_F$ and $\Omega\sim\epsilon T$. Therefore in the leading order in $\epsilon$, the blocks $f_n^{R}(2E-\Omega,q)$ in Eqs.~\eqref{SR-clean} and \eqref{SZ-clean} can be evaluated at zero momentum and frequency: $f_n^{R}(2E-\Omega,q)\approx(1\mp2 i \tau E)^{-n}$. This leads to an additional suppression of both $\Sigma^R$ and $\Sigma^Z_\text{tanh}$, which now behave as $\Sigma^R\sim\Sigma^Z_\text{tanh}/\Omega\sim{\cal{O}}(1/\tau^3)$. The corresponding correction to conductivity then becomes $\tau$-independent.

It is instructive to trace this cancellation in terms of particular diagrammatic contributions:
\begin{subequations}
\label{sigma-ball-cancellation}
\begin{align}
\label{sigma-ball-dos}
\delta\sigma^\text{DOS}
& =
\left[-\frac{2\pi^2}{7\zeta(3)}(T\tau)^2-\frac{2}{\pi}T\tau + {\cal{O}}(1)\right]\ln\frac{1}{\epsilon},
\\
\label{sigma-ball-rmt}
\delta\sigma^\text{MT}_\text{tanh}
& =
\left[\frac{2\pi^2}{7\zeta(3)}(T\tau)^2-\frac{2}{\pi}T\tau + {\cal{O}}(1)\right]\ln\frac{1}{\epsilon},
\\
\label{sigma-ball-al}
\delta\sigma^\text{AL1}
& =
\left[\frac{4}{\pi}T\tau + {\cal{O}}(1)\right]\ln\frac{1}{\epsilon} .
\end{align}
\end{subequations}
Here cancellation of the quadratic terms $(T\tau)^2$ is a general feature of the ballistic correction discussed in Sec.\ \ref{SS:ball-gen}, whereas cancellation of the linear terms $T\tau$ takes place only for the leading $\ln1/\epsilon$ contribution in the vicinity of the transition. 
The remaining $\delta\sigma_\text{tanh} = {\cal{O}}(1)\ln(1/\epsilon)$
will be discussed in Sec.\ \ref{SS:cross-tanh}.

Cancellation of the linear in $T\tau$ terms in Eqs.\ (\ref{sigma-ball-cancellation}) forces us to analyze the singular part of the AMT contribution.
The function $\Sigma^Z_\text{sing}(\Omega,q)$ behaves differently for small and large momenta. In the diffusive limit ($ql\ll1$), only the poles of cooperons $\gamma^{R,A}$ are important. In the ballistic limit ($ql\gg1$), it is given by Eq.~(\ref{SZ-clean-sing-res}). Thus we find
\begin{equation}
\label{Sigma-Z-sing}
\Sigma^Z_\text{sing}(\Omega,q)
\approx
-\frac{\Omega}{2T\tau^2}
\begin{cases}
1/2Dq^2, & ql\ll 1, 
\\
{\ln(qv_F\tau)}/{\pi qv_F}, & ql\gg 1.
\end{cases}
\end{equation}
As mentioned above, the form of the propagator \eqref{LA-ball-near} sets the temperature-dependent coherence length $\xi_c/\sqrt\epsilon$ as the relevant spatial scale. Depending on its relation to $l$ one should distinguish between the two limiting regimes: the \emph{moderately clean}\ case, $\xi_c\ll l\ll\xi_c/\sqrt{\epsilon}$ [region (c$'$)], and the \emph{superclean}\ case, $\xi_c/\sqrt{\epsilon}\ll l$ [region (c$''$) in Fig.~\ref{fig:phasediagram}].

\begin{itemize}

\item
In the \emph{moderately clean} case, the relevant momenta belong to the diffusive region, $q l\ll 1$. In this regime we can simply use the diffusive result for the AMT contribution \eqref{sigma-AMT-res-diff-close}, 
provided the diffusive fluctuation propagator (\ref{G(z)}) is replaced by the ballistic one, Eq.~(\ref{LA-ball-near}).
Hence we obtain \cite{Maki1990}
\begin{equation}
\label{AMT-ball-near-clean}
\delta\sigma^\text{AMT}_\text{sing, mod.cl.}
\approx
\frac{1}{8}\frac{1}{\epsilon-\gamma'}\ln\frac{\epsilon}{\gamma'},
\end{equation}
where $\gamma'^{-1} = [16\pi^2/7\zeta(3)] (T\tau) (T\tau_\phi)$. The parameter $\gamma'$ is related to its diffusive counterpart $\gamma$ 
via $\gamma'/\gamma = (\xi_c/\xi_d)^2$, see Sec.~\ref{SS:5cut}.

\item
In the \emph{superclean} case, one should take both the contributions from the diffusive and ballistic momenta into account. The diffusive contribution is given by Eq.~(\ref{AMT-ball-near-clean}), whereas the ballistic contribution is calculated using Eqs.\ (\ref{ds-via-Sigma-ball}) and (\ref{Sigma-Z-sing}):
\begin{equation}
\label{AMT-ball-near-supclean}
\delta\sigma^\text{AMT}_\text{sing, sup.cl.}
\approx
\delta\sigma^\text{AMT}_\text{sing, mod.cl.}+
\frac{\pi T\tau}{\sqrt{14\zeta(3)\epsilon}}
\ln\frac{l\sqrt\epsilon}{\xi_c}.
\end{equation}
The second term in Eq.~\eqref{AMT-ball-near-supclean} has been derived in Ref.~\cite{Randeria} (with the coefficient $\pi$ times larger due to an arithmetical error). Due to the logarithmic factor, it predicts that the AMT correction grows faster with $\tau$ than the Drude conductivity.
In fact, Eq.~\eqref{AMT-ball-near-supclean} is valid as long as the mean free path $l$ is shorter than the inelastic length $l_\text{in}$. In the opposite case, $l_\text{in}<l$,
the logarithm should be replaced by $\ln(l_\text{in}\sqrt\epsilon/\xi_c)$, restoring the scaling $\delta\sigma^\text{AMT}_\text{sing, sup.cl.} \propto \tau$ for very clean samples.

\end{itemize}

Finally, there is the AL2 term (\ref{AL2}), whose leading contribution is insensitive to disorder near $T_c$ and is given by Eq.\ \cite{AL}.
Though it remains finite in the limit $T\tau\to\infty$, it should be retained as it is more singular near the transition and thus competes with the AMT contribution.

Hence we see that in the ballistic limit in the vicinity of $T_c$ the fluctuation correction to conductivity is determined by the interplay of the AMT [Eqs.\ (\ref{AMT-ball-near-clean}) or (\ref{AMT-ball-near-supclean})] and AL2 [Eq.~\eqref{AL2}] contributions. The former contains a large factor $T\tau$, while the latter has a stronger singularity at $\epsilon\to0$.

Our results are in contradiction with the conclusion of Ref.~\cite{Dorin93}, where the ballistic correction was claimed to be proportional to $(T\tau)^2$ in the moderately clean limit. This strongly growing with $\tau$ fluctuation correction was attributed to the DOS diagram, in accordance with our result (\ref{sigma-ball-dos}). However we find that the pole contribution of the MT diagram has the same leading behavior but with the opposite sign [see Eq.~\eqref{sigma-ball-rmt}], that leads to the exact compensation of the $(T\tau)^2$ terms
(as explained in Sec.\ \ref{SS:ball-gen}, this cancellation is general, holding not only in the vicinity of $T_c$). Thus we believe that the discrepancy with Ref.~\cite{Dorin93} is due to their incorrect evaluation of the MT diagram. Understanding that the growth of $\delta\sigma$ as fast as $(T\tau)^2$ is unphysical, in Ref.~\cite{Livanov2000} 
it was suggested (and even numerically confirmed) that this behavior is realized only in the moderately clean case, while in the superclean case `the large negative DOS contribution can be cancelled with the positive anomalous MT one'.
However our analysis demonstrates that this explanation is flawful: the cancellation of the most divergent terms takes place already in the moderately clean limit due to the pole contribution of the MT diagram.

\subsection{Far above the transition, $T\gg T_c$}

\subsubsection{\textbf{Ultraviolet divergency}}
\label{SSS:uv-div}

At high temperatures [region (d) in Fig.~\ref{fig:phasediagram}], the leading contribution to the fluctuation correction is determined by the term with $b(y)=B(\Omega)$ in Eq.~(\ref{ds-via-Sigma-ball}), where both frequency and momentum can be large: $\Omega$, $qv \gg T_c$. In this region, $L^R$ is a logarithmically slow function, while $\Sigma^R$ given by Eq.~(\ref{SR-clean}) can be roughly estimated as $1/[\tau^2\max^2(\Omega,qv)]$, and with logarithmic accuracy we obtain
\be
\label{divergent}
\delta\sigma
\sim 
v^2 \tau
\int 
\frac{d\Omega \, d^dq}{\max^2(\Omega,qv)}
\sim 
v \tau
\int
\frac{d^dq}{q}
\sim 
v \tau
q_\text{max}^{d-1} .
\ee
The correction
diverges in the ultraviolet for all space dimensionalities $d\geq1$ [for $d=1$, $\delta\sigma\propto\ln\ln q_\text{max}$ due to the logarithmic dependence of the fluctuation propagator omitted in Eq.~(\ref{divergent})].
In the ballistic limit close to $T_c$, the blocks $\Sigma^{R,Z}$ can be evaluated at zero $\Omega$ and $q$, and convergence of the integrals for the fluctuation correction is provided by $L^R$ [see Eq.\ \eqref{LA-ball-near}]. In the ballistic region at $T\gg T_c$, the situation is opposite: here $L^R$ is nearly constant, and a weak decay of $\Sigma^R(\Omega,q)$ cannot compensate the growing phase volume of superconducting fluctuations with large frequency and momentum.

Such a behavior should be compared with the situation at $T\gg T_c$ in the diffusive regime [region (b) in Fig.~\ref{fig:phasediagram}], where the presence of diffusive poles in $\Sigma^R$ and $\Sigma^Z$ provides a better convergence of the integrals in Eq.~(\ref{ds-via-Sigma}). Nevertheless, for $d\geq2$ the fluctuation correction in the diffusive case still diverges in the ultraviolet $\propto q_\text{max}^{d-2}$, where $q_\text{max}\sim1/l$ is the upper boundary of the diffusive limit. It is this divergency that leads to the $\ln\ln1/T_c\tau$ term in Eq.~\eqref{sigma-reg-diff-far}.
In the ballistic region, it transforms to a stronger divergency in Eq.\ \eqref{divergent}.
Strong temperature-independent infrared divergency in the ballistic limit was first discussed by Zala \emph{et al.} \cite{Zala}.

Three interrelated properties of the fluctuation propagator simplify the analysis at large temperatures ($T\gg T_c$): (i) its smallness, $L^R<1/\eps\ll1$, (ii) its logarithmically weak dependence on $\Omega$ and $q$, and (iii) the fact that $\Im L^R \ll \Re L^R$.

\subsubsection{\textbf{Contribution of $\Sigma^R$}}
\label{SSS:clean-far-eval-R}

Consider first the contribution of $\Sigma^R$ to Eq. \eqref{ds-via-Sigma-ball}, that will be denoted as $\delta\sigma^R$.
Neglecting $\Im L^R$ for the reason described above and taking $\Sigma^R$ from Eq.~(\ref{SR-clean-res}) we get
\begin{multline}
\delta\sigma^R
\approx
\frac{2 T\tau}{\pi}
\int_{-\infty}^\infty dy
\int_0^\infty x \, dx
\coth(2\pi y) L^R 
\\{}
\times
\int_0^\infty dt \,
\frac{t^2 [J_0(xt) - S_0(xt)] \sin(yt)}{2\sinh t/2}
,
\end{multline}
where $J_0(z)$ is the Bessel function, and $S_0(z)=\sin(z)/z$.
Next we integrate over $y$ neglecting a logarithmically weak dependence of $L^R$ on $y$:
\be
\label{ds-ball-far-1}
\delta\sigma^R
=
\frac{4T\tau}{\pi}
\int_0^\infty x \, dx \, L^R
\int_0^\infty dt \,
\frac{t^2 [J_0(xt) - S_0(xt)]}{16\sinh^2t/4} .
\ee

As already discussed in Sec.\ \ref{SSS:uv-div}, this integral diverges in the limit $x\to\infty$. 
Since large $x$ correspond to small $t$, one can replace $(t/4)^2/\sinh^2(t/4)$ by 1, and the integral over $t$ gives $(1 - \pi/2)/x$. Then cutting the remaining integral over $x$ at $x=x_\text{max}\gg1$ we obtain
\be
\label{ds-div-ball}
\delta\sigma_\text{div}^R
\sim
- T\tau
\frac{x_\text{max}}{\epsilon+\ln x_\text{max}}
\sim
- \frac{v\tau q_\text{max}}{\ln(vq_\text{max}/T_c)} ,
\ee
in accordance with the estimate \eqref{divergent}.
The key point is that this divergent correction is temperature independent and hence can be absorbed into the Drude conductivity
\cite{Zala}.

The temperature-dependent contribution to the fluctuation conductivity is given by the subleading term in Eq.~\eqref{ds-ball-far-1} originating from $x\sim t\sim 1$ (and hence $y\sim1$). 
In order to find it, we should account for the difference between $(t/4)^2/\sinh^2(t/4)$ and 1 in Eq.\ \eqref{ds-ball-far-1}. Since the resulting integrals converge sufficiently fast, we may replace $L^R$ by $1/\epsilon$. Then integrating over $x$ as $\int_0^\infty dx \,x \,[J_0(xt) - S_0(xt)]=\delta(t)/t - 1/t^2$, we get
\be
\delta\sigma^R_T
=
\frac{4T\tau}{\pi\eps}
\int_0^\infty \frac{dt}{t^2}
\left[ 1 - \frac{(t/4)^2}{\sinh^2t/4} \right] .
\ee
The final integration gives the leading $T$-dependent correction:
\be
\label{ds-ball-RT}
\delta\sigma^R_T
=
\frac{T\tau}{\pi\eps} .
\ee
A more accurate estimate of the integrals as well as account for $\Im L^R$ would generate smaller terms of the order of $T\tau\,{\cal O}(1/\epsilon^2)$.

The interaction correction to the conductivity of clean normal metals has been studied in Ref.~\cite{Zala}.
In the case of weak point-like interaction with the dimensionless interaction amplitude $\lambda=\nu V_0$, their Eq.\ (2.14) predicts $\delta\sigma_\text{ZNA}=-\lambda T\tau/\pi$.
The case of short-range attractive interaction studied in our paper formally corresponds to $\lambda=-1/\ln(T/T_c)$. Substituting such $\lambda$ into $\delta\sigma_\text{ZNA}$, one immediately recovers our expression (\ref{ds-ball-RT}).

\subsubsection{\textbf{Contribution of $\Sigma^Z$}}
\label{SSS:clean-far-eval-Z}

Now we turn to the analysis of the contribution from $\Sigma^Z=\Sigma^Z_\text{tanh}+\Sigma^Z_\text{sing}$ to Eq.~\eqref{ds-via-Sigma-ball}, that will be denoted as $\delta\sigma^Z=\delta\sigma^Z_\text{tanh}+\delta\sigma^Z_\text{sing}$. Due to the presence of $b'(y)$ this correction does not contain a divergent part, and its high-temperature expansion contains $1/\epsilon^2$ originating from $\Im L^R$. So one could expect that $\delta\sigma^Z$ is always smaller than $\delta\sigma^R$. However, this is not the case due to the presence of a special cut contribution $\Sigma^Z_\text{sing}$ from the AMT diagram, which does not appear in $\Sigma^R$. 

Calculation of $\delta\sigma^Z_\text{tanh}$ and $\delta\sigma^Z_\text{sing}$ is quite straightforward. Using Eqs.~\eqref{ds-via-Sigma-ball}, \eqref{LR-inv-ball-res}, \eqref{SZ-clean-sing-res} and \eqref{SZ-clean-poles-res}, we obtain
\begin{gather}
\label{ds-ball-Z-poles}
\delta\sigma^Z_\text{tanh}
=
\alpha_p
\frac{T\tau}{\epsilon^2} ,
\\
\label{ds-ball-Z-sing}
\delta\sigma^Z_\text{sing}
=
\alpha_c \frac{T\tau \ln T\tau}{\eps^2} ,
\end{gather}
where $\alpha_p=0.458$ and $\alpha_c=0.190$ are the numeric values for the integrals (\ref{int-ap}) and (\ref{int-ac}).

\subsubsection{\textbf{Resulting expression}}

Taking into account Eqs.\ \eqref{ds-ball-RT}, \eqref{ds-ball-Z-poles} and \eqref{ds-ball-Z-sing}, we obtain for the leading asymptotic behavior: 
\be
\label{sigma-tot-ball-far}
\delta\sigma 
= 
\const
+
\frac{T\tau}{\pi\eps} 
+
0.190
\frac{T\tau \ln T\tau}{\eps^2} + \dots
\ee
Keeping the term (\ref{ds-ball-Z-poles}) in the resulting expression is beyond the accuracy, as it is of the same order as the subleading term from Eq.\ (\ref{ds-ball-RT}) also proportional $1/\epsilon^2$. However the AMT term (\ref{ds-ball-Z-sing}) is retained since it contains an additional large factor of $\ln T\tau$. As it was discussed in Sec.~\ref{SS:clean-close}, the logarithmic factor in the AMT contribution should be replaced by $\ln T\tau_\text{in}$ as long as the mean free time $\tau$ exceeds the inelastic scattering time $\tau_\text{in}$ due to interaction (very clean limit).

Fluctuation conductivity in the absolutely clean limit (no impurities, $\tau=\infty$) was studied in Ref.\ \cite{Reggiani1991}. The limit $\tau\to\infty$ was taken first while keeping $\omega$ finite, and then $\omega$ was set to zero. It was reported that the sum of the DOS and MT corrections was proportional to $\omega$ and thus did not contribute to the dc conductivity. In the limit $T\gg T_c$, the remaining AL contribution was claimed to scale as $1/\epsilon^3$. On the contrary, we first take the limit $\omega=0$ and then arrive at a much larger expression (\ref{sigma-tot-ball-far}), which grows roughly as $T\tau$.
The discrepancy between our result and that of Ref.\ \cite{Reggiani1991} indicates that the limits $\omega\to0$ and $\tau\to\infty$ do not commute. Since some amount of disorder is inevitably present in real samples, we argue that in calculating the dc conductivity the limit $\omega\to0$ should be taken first. Frequency dependence of the fluctuation correction in the ballistic limit will be calculated elsewhere \cite{2B}.

\section{Dirty/Clean crossover near $T_c$}
\label{S:cross}

This Section reviews the behavior of the fluctuation correction in the vicinity of the transition ($T\to T_c$) for an arbitrary disorder strength characterized by the parameter $T\tau$. The resulting expressions will describe the crossover between the dirty [region (a) in Fig.\ \ref{fig:phasediagram} discussed in Sec.\ \ref{SS:diff-close}] and clean [region (c) discussed in Sec.\ \ref{SS:clean-close}] limits.
We will follow the approach of Sec.~\ref{SS:ball-gen} and separate the AMT correction (\ref{S_Z-AMT}) into the contributions of the singularities of $f_1^A\gamma^A\gamma^R$ ($\delta\sigma^\text{AMT}_\text{sing}$) and the poles of the distribution function ($\delta\sigma^\text{AMT}_\text{tanh}$). 
Hence we have three contributions: (i) $\delta\sigma^\text{AL2}$, (ii) $\delta\sigma_\text{tanh}$ due to the poles of $F_E$ (which includes DOS, AL1, RMT and a part of MT) and (iii) $\delta\sigma^\text{AMT}_\text{sing}$ due to cooperons in AMT.

\subsection{AL2 contribution}

The two-propagator AL2 contribution (\ref{sigma-AL2-v2}) gives the leading Aslamazov-Larkin correction $\delta\sigma^\text{AL2}\approx{1}/{16\epsilon}$, which is known to be insensitive to disorder \cite{AL}. However less-singular subleading terms 
proportional to $\ln(1/\epsilon)$ 
do depend on disorder.
To find them we have to calculate the fluctuation propagator with a higher precision.
Retaining the next-order terms
$(qv_F)^4$, $\Omega^2$ and $(qv_F)^2\Omega$ 
in Eq.\ \eqref{LR-cross-near}, we get after some algebra:
\be
\delta\sigma^\text{AL2}=\frac{1}{16\epsilon}+
s(T\tau)
\ln\frac{1}{\epsilon}+{\cal{O}}(\epsilon^0) .
\ee
Here the crossover function $s(T\tau)$ can be expressed in terms of the function ${\cal F}$ introduced in Eq.~\eqref{F-def} as follows:
\begin{multline}
\label{function-s-def}
s(T\tau)=
\frac{\psi''(1/2)}{4\pi^{4}}
-
\frac{1}{4\pi^{2}}\frac{{\cal F}'(1/2)}{{\cal F}(1/2)}+{}
\\{}
+
\frac{6{\cal F}(1/2)+[2\psi''(1/2)+\psi''(1/2+1/4\pi T\tau)]}
{32\,{\cal F}^{2}(1/2) (4\pi T\tau)^{2}} .
\end{multline}
This is a monotonous function interpolating from $s(0)=0$ in the diffusion limit [in accordance with Eq.~\eqref{ccc}] to $s(\infty)\approx0.0183$ in the ballistic limit.

\subsection{Poles of $F_E$ contribution}
\label{SS:cross-tanh}

The contribution of the poles of $F_E$ in Eqs.\ (\ref{S_R-DOS+RMT}), (\ref{S_Z-DOS+RMT}), (\ref{S_Z-AMT}) and (\ref{S_R-AL1}) is proportional to $\ln(1/\epsilon)$. The corrections with an additional cooperon, Eqs.\ (\ref{S_R-MT(C)})--(\ref{S_Z-DOS(C)}), are not singular near $T_c$ and will be disregarded. 
Evaluating the functions (\ref{fg}) in the first non-vanishing order in $\Omega$ and $q$, and summing the contributions of the poles of $F_E$ in the doing the integrals over $E$ in the blocks $\Sigma^Z_\text{tanh}$ and $\Sigma^R$, we arrive at surprisingly concise expressions:
\begin{equation}
\label{S_Z-S_R-cross}
\Sigma^Z_\text{tanh}(\Omega,q)
=
-\frac{\Omega\, {\cal{F}}'(1/2)}{\pi^2T\tau},
\qquad
\Sigma^R(\Omega,q)
=0.
\end{equation}
Using then Eq.\ (\ref{ds-via-Sigma}) with the local form (\ref{LR-cross-near}) of the fluctuation propagator, we obtain for the leading conductivity correction:
\be
\delta\sigma_\text{tanh}=
\frac{1}{2\pi^2}\frac{{\cal{F}}'(1/2)}{{\cal{F}}(1/2)}\ln\frac{1}{\epsilon}
+{\cal{O}}(\epsilon^0).
\ee
This crossover formula interpolates between the diffusive and ballistic regimes:
\be
\label{ds-tanh-limiting}
\delta\sigma_\text{tanh}
=
\ln\frac{1}{\epsilon} \times
\begin{cases}
- {14\zeta(3)}/{\pi^4}, & T\tau\ll 1,
\vspace{2mm}
\\
- {\pi^2}/{28\zeta(3)}, & T\tau\gg 1.
\end{cases}
\ee
In the notations of Sec.\ \ref{SS:diff-close}, the diffusive limit of Eq.~\eqref{ds-tanh-limiting} corresponds to $c_\text{tanh}=-14$, which differs from $c^\text{REG}=-7$ in Eq.~(\ref{ccc}).
The reason for this discrepancy is that unlike $\sigma_\text{tanh}$, $\sigma^\text{REG}$ does not contain a contribution coming from the poles of $F_E$ in AMT term.

\subsection{Singularities of the AMT contribution}
\label{SS:5cut}

The analysis of this term has already been performed in Sec.~\ref{SS:clean-close}. The only difference is in the form the fluctuation propagator \eqref{LR-cross-near} which contains the disorder-dependent coherence length $\xi(T\tau)$ defined in Eq.~\eqref{xi-def}. 

Therefore we conclude that both in the \emph{diffusive} and \emph{moderately clean} cases (when $T\tau\ll 1/\sqrt{\epsilon}$) one can use the result (\ref{AMT-ball-near-clean}), provided $\gamma'$ is replaced by $\gamma(T\tau)$:
\begin{equation}
\label{gamma-cross-def}
\gamma(T\tau)=\frac{\xi^{2}(T\tau)}{D\tau_{\phi}}={\cal{F}}(1/2)\frac{\tau}{\tau_{\phi}} .
\end{equation}
This function interpolates between $\gamma$ in the diffusive regime and $\gamma'$ in the ballistic regime.

In the \emph{superclean} case ($T\tau\gg1/\sqrt{\epsilon}$), there is an additional contribution from high momenta given by the second term in Eq.\ (\ref{AMT-ball-near-supclean}).

\section{Discussion and conclusion}
\label{S:Discussion}

In this work, we have studied the Cooper-channel contribution to the fluctuation conductivity of disordered $s$-wave superconductors above the transition temperature. Working in the leading, one-loop approximation in the fluctuation propagator, we have derived the general expression for the fluctuation correction valid for arbitrary temperatures $T>T_c$ and disorder strengths characterized by the parameter $T\tau$. The result obtained for an arbitrary space dimensionality is then analyzed in the experimentally relevant 2D case.

Our approach is based on the usual diagrammatic technique \cite{LV-book}, yet in the Keldysh representation. 
The use of the Keldysh technique was crucial for proper disorder averaging in blocks of electron Green's functions.
This operation is sensitive to a particular combination of retarded and advanced Green's functions in the block, and each particular combination should be treated in a different manner. The advantage of the Keldysh technique is that all the blocks made of $G^R$ and $G^A$ are generated automatically, whereas in the Matsubara technique that requires a tedious procedure of analytic continuation, which might be an additional source of possible computational errors. Though both approaches are in principle equivalent, the simplicity of the Keldysh machinery becomes especially beneficial beyond the diffusive limit, where all possible combinations of $G^R$ and $G^A$ should be taken into account.

Our analysis demonstrates that the standard classification of the diagrammatic contributions involving the AL, DOS and MT terms does not unambiguously reflect the underlying physical processes.
We see that the resulting expressions for different disorder-averaged diagrams contain a number of similar fragments. Therefore it is natural to split them 
and rearrange the terms according to their mathematical structure. In particular, we split the AL contribution into AL1 (one fluctuation propagator) and AL2 (two fluctuation propagators), with the AL1 term partially cancelling the contribution of the DOS and MT diagrams. This illustrates the fact that each of the individual AL, DOS and MT diagrams can hardly be ascribed a well-defined physical meaning. It is the sum of all diagrams that contains the effects of paraconductivity, DOS suppression and scattering on superconducting fluctuations, and these physical processes cannot be uniquely associated with the standard AL, DOS and MT diagrams. This circumstance has been recognized in a number of recent publications \cite{Glatz2011,Tikhonov,Petkovic}, where different classification schemes have been suggested.

The derived analytical expression is used to critically revise previous results for the fluctuation correction. The majority of these results was obtained with the help of the Matsubara diagrammatic technique and are available only in the four asymptotic regions (close to $T_c$/far above $T_c$ and in the diffusive/ballistic limit) shown in Fig.~\ref{fig:phasediagram}.

In the diffusive limit ($T\tau\ll1$), the full temperature dependence has been recently addressed in Refs.\ \cite{Glatz2011,Tikhonov,Petkovic}.
We entirely reproduce the results of Refs.\ \cite{Tikhonov,Petkovic} for the fluctuation correction at arbitrary temperatures. 
Notably the same result has been obtained by three independent groups that used different versions of the Keldysh technique.

In the ballistic limit ($T\tau\gg1$), we provide the first analytical description of the fluctuation correction at arbitrary temperatures. Here we observe a significant discrepancy with previous calculations obtained within the Matsubara technique. Our main conclusions can be summarized as follows:
\begin{itemize}
\item
In the vicinity of $T_c$ [region (c) in Fig.~\ref{fig:phasediagram}], it is only the anomalous part of the MT correction that behaves differently in the moderately clean (c$'$) and superclean (c$''$) limits, see Eqs.\ \eqref{AMT-ball-near-clean} and \eqref{AMT-ball-near-supclean}. All other contributions are insensitive to the moderately clean/superclean crossover.
In this region we report the absence of the terms growing as $(T\tau)^2$, contrary to the claims of Refs.\ \cite{Dorin93,Livanov2000}.
\item
At high temperatures ($T\gg T_c$), we find that the fluctuation correction contains a formally divergent temperature-independent term that should be incorporated into the bare conductivity. The remaining temperature-dependent correction grows as $T\tau$ up to some logarithmic factors, see Eq.~\eqref{sigma-tot-ball-far}. This result is consistent with Ref.~\cite{Zala} and contradicts the $1/\epsilon^3$ decay reported in Ref.\ \cite{Reggiani1991}, where the limit $\tau\to\infty$ was taken before the limit $\omega\to0$.
\end{itemize}

\begin{figure}
\includegraphics[width=0.96\columnwidth]{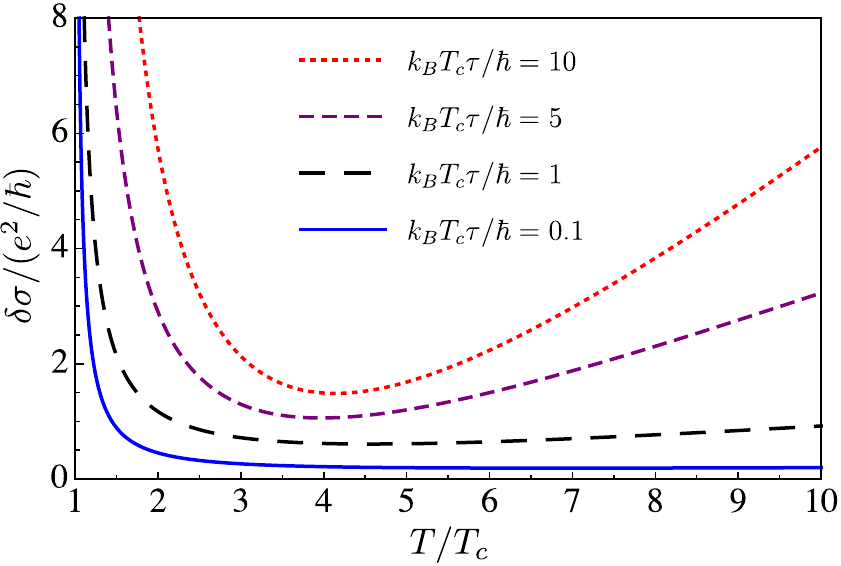}\hspace{0.02\columnwidth}
\caption{
Temperature dependence of the total superconducting correction to the conductivity of a 2D film in units of $e^2/\hbar$ above the transition. The curves correspond to different disorder strength measured by the parameter $k_BT_c\tau/\hbar$: $0.1$ (lower curve, diffusive case), 1, 5, and 10 (upper curve, ballistic case). The dephasing time is taken to be $\tau_\phi=100\hbar/k_BT_c$ for all disorder strengths. The curves are vertically offset for clarity.}
\label{fig:total_picture}
\end{figure}

In Fig.~\ref{fig:total_picture}  we present the temperature dependence of the total fluctuation correction to the conductivity of a 2D film obtained numerically for different disorder strengths (here we recover the physical units).
As the correction diverges in the ultraviolet (see discussion in Sec.\ \ref{SSS:uv-div}), $\delta\sigma$ is determined up to a temperature-independent constant chosen arbitrarily in Fig.~\ref{fig:total_picture}.

Two features of the temperature dependence of $\delta\sigma$ are to be pointed out.
First, we see that near the transition the correction grows with $\tau$. Since Fig.~\ref{fig:total_picture} is plotted for a week dephasing, $\gamma\approx0.004$, the visible parts of the curves at $T\to T_c$ are described by the AMT contribution rather than by the universal AL contribution. In the diffusive limit, the AMT correction is insensitive to $\tau$, whereas in the ballistic limit it grows with $\tau$ according to Eqs.\ \eqref{AMT-ball-near-clean} and \eqref{AMT-ball-near-supclean}.
Second, far above the transition the fluctuation correction grows with $T$ for all disorder strengths as predicted by Eqs.~(\ref{sigma-reg-diff-far}) and (\ref{sigma-tot-ball-far}). Even for $k_BT_c\tau/\hbar=0.1$ (the most diffusive sample), $d\delta\sigma/dT>0$ at $T/T_c=10$. Thus the dependence $\delta\sigma(T)$ has a minimum for all disorder strengths.
Note that this minimum (corresponding to the maximum in the temperature dependence of the resistance) is obtained even in the absence of the weak localization and interaction corrections!

Knowledge of the exact expression for the temperature dependence of the fluctuation conductivity is not only of academic interest. In combination with the weak-localization and interaction corrections \cite{AA,Zala}, it provides a powerful tool for the high precision determination of $T_c$ from transport measurements \cite{SacepeBaturina2010}. Extension of the theory to the case of arbitrary disorder achieved in our work opens a way for application of the same technique to clean superconductors (e.g., NbSe$_2$ \cite{NbSe2} or organic superconductors \cite{organic-book,organic-dis}). 

As an open question we would like to mention the sensitivity of the fluctuation correction to microscopic details of the disorder potential that become important in the ballistic limit. In our analysis, we assumed the simplest form of a weak Gaussian disorder [see Eq.~(\ref{<UU>})]. It is a natural model in the diffusive limit, where the observables are expressed only in terms of the diffusion coefficient $D$ and do not involve the mean-free time $\tau$ itself. In the ballistic limit, the result is sensitive to a particular type of disorder and might be different for the white-noise and Poissonian statistics of the random potential \cite{strong-imp,FS2016}. This concern applies to all previous studies of fluctuation corrections to conductivity in the ballistic limit and to other superconducting fluctuation phenomena in this regime (see, e.g., Ref.\ \cite{Levchenko-tun-bal}).

Finally, we mention an interesting direction of generalizing the developed theory for the fluctuation conductivity to the case of unconventional superconductors. In these materials nonmagnetic disorder is known to acts as an effective pair breaker \cite{Larkin65}, such that the system is always in the clean limit.

\acknowledgments

We are grateful to I. S. Burmistrov, M. V. Feigel'man, P. M. Ostrovsky, and K. S. Tikhonov for stimulating discussions. 
The research was partially supported
by Skoltech NGP Program (Skoltech-MIT joint project)
and by the Russian Science Foundation (Grant No.\ 14-42-00044).

\begin{figure*}
\includegraphics[width=1.9\columnwidth]{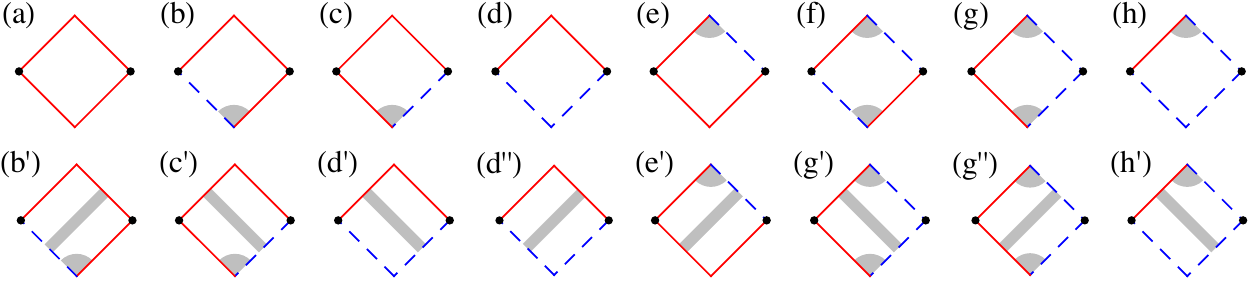}
\caption{Disorder averaging of various blocks that appear in calculating the MT diagram shown in Fig.~\ref{fig:Diagrams}. Solid red lines denote $G^R$ and dashed blue lines denote $G^A$, cooperons are shown in gray, 
and the current vertices are depicted by dots.
There are similar diagrams with $G^R\leftrightarrow G^A$.
The second line shows the results of averaging with an additional cooperon, denoted as MT(2).
In the diffusive regime, only the diagrams with the largest number of cooperons [(f), (g), (g$'$) and (g$''$)] are important. The diagram (f) is responsible for the anomalous MT contribution.
}
\label{fig:Blocks-MT}
\end{figure*}
\begin{figure}[b]
\includegraphics[width=0.5\columnwidth]{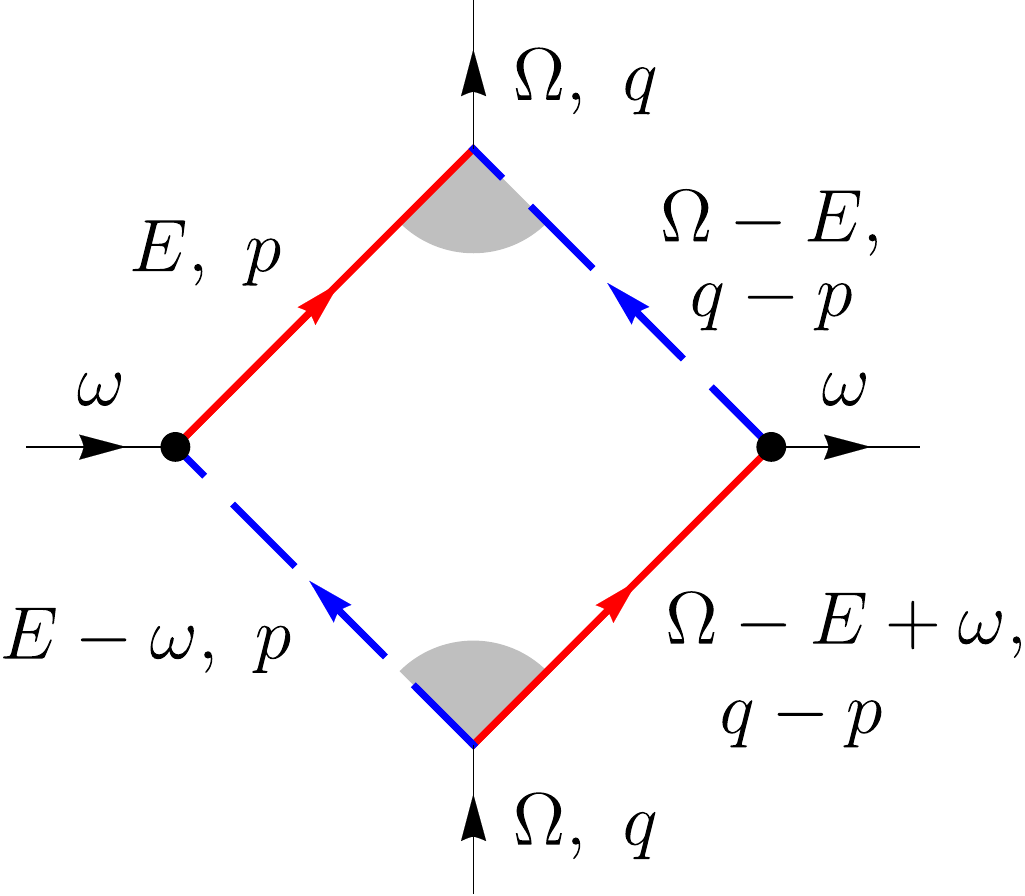}
\caption{The AMT diagram shown in Fig.~\ref{fig:Blocks-MT}(f) with the energy 
and momentum arguments indicated.}
\label{fig:Block-AMT}
\end{figure}

\appendix

\section{Initial expression for the diagrams}
\label{A:Initial expression}

The analytic expressions for the diagrams shown in Fig.~\ref{fig:Diagrams} have the form:
\begin{multline}
\label{MAS-QDOS-ini}
Q^{\text{DOS}}(\omega)=\frac{i}{2}\int\frac{dE}{2\pi}\int\frac{d\Omega}{2\pi}\int d\br_1d\br_2d\br_3
\\{}
\times\mathrm{tr}\big\{
\sigma_1\hat{v}_xG_E(\br_0,\br_1)\gamma^a\tilde G_{E-\Omega}(\br_1,\br_2)
\\{}
\times
\gamma^bG_E(\br_2,\br_3)\hat{v}_xG_{E-\omega}(\br_3,\br_0)
\\{}
+
\sigma_1\hat{v}_xG_{E}(\br_0,\br_3)\hat{v}_x G_{E-\omega}(\br_3,\br_1)
\\{}
\times
\gamma^a\tilde G_{E-\Omega-\omega}(\br_1,\br_2)\gamma^b G_{E-\omega}(\br_2,\br_0)
\big\}
\\{}
\times L_{ab}(\Omega;\br_1-\br_2)
,
\end{multline}
\vskip-6mm
\begin{multline}
\label{MAS-QMT-ini}
Q^{\text{MT}}(\omega)=
-
\frac{i}{2}\int\frac{dE}{2\pi}\int\frac{d\Omega}{2\pi}\int d\br_1d\br_2d\br_3
\\
\times\tr\big\{
\sigma_1\hat{v}_xG_E(\br_0,\br_1)\gamma^a\tilde G_{E-\Omega}(\br_1,\br_2)\hat{v}_x\tilde G_{E-\Omega-\omega}(\br_2,\br_3)
\\
\gamma^b G_{E-\omega}(\br_3,\br_0)
\big\}L_{ab}(\Omega;\br_1-\br_3),
\end{multline}
\vskip-6mm
\begin{multline}
\label{MAS-QAL-ini}
Q^{\text{AL}}(\omega)=-2\left(\frac{i}{2}\right)^2\int\frac{dE}{2\pi}\int\frac{dE^{\prime}}{2\pi}\int\frac{d\Omega}{2\pi}\int d\br_1\dots d\br_5
\\
\times\tr\big\{
\sigma_1\hat{v}_xG_E(\br_0,\br_1)\gamma^a\tilde G_{E-\Omega}(\br_1,\br_2)\gamma^b G_{E-\omega}(\br_2,\br_0)\big\}
\\
\times\tr\big\{
\hat{v}_x G_{E^{\prime}-\omega}(\br_3,\br_4)\gamma^c\tilde G_{E^\prime-\Omega}(\br_4,\br_5)\gamma^d G_{E^\prime}(\br_5,\br_3)\big\}
\\
\times L_{ad}(\Omega;\br_1-\br_5)L_{cb}(\Omega-\omega;\br_4-\br_2).
\end{multline}
Here $\tr$ acts only in the Keldysh space as we have already taken the trace in the Nambu space employing the diagonal form of the Green's function (\ref{G-diag-N}).

Expressions for the kernels (\ref{MAS-QDOS-ini})--(\ref{MAS-QAL-ini}) should be averaged over disorder. 
Using Eqs.\ (\ref{G-K-all}) and taking the trace over the Keldysh space, one obtains a product of several $G^R$ and $G^A$ that should be averaged with the standard impurity diagrammatic technique.

\section{Disorder averaging}
\label{A:Disorder averaging}

In this Section we illustrate how the general procedure of disorder averaging outlined in Appendix \ref{A:Initial expression} is implemented for the MT contribution. 
Calculating the trace in Eq.~(\ref{MAS-QMT-ini}) one encounters $2^4=16$ blocks (labeled by $n$) with four lines, where each Green's function can be either $G^R$ or $G^A$.
Averaging them independently from the fluctuation propagator, we reduce the kernel to the sum of partial contributions of the form
\begin{equation}
\label{eq:Qf}
Q_{(n)}^\text{MT}(\omega)=\frac{i}{2}\int(d\Omega)(d\mathbf{q})(dE)
\,{\cal{B}}^\text{MT}_{(n)}{\cal{L}}^\text{MT}_{(n)},
\end{equation}
where ${\cal{B}}_n$ is the averaged block, while ${\cal{L}}_n$ contains both the propagator and the distribution function [arising from $G^K$ in Eqs.\ (\ref{G-K-all})].
Hereafter we use the notations $(d\Omega)\equiv d\Omega/2\pi$ and $(d\mathbf{q})=d^d{q}/(2\pi)^d$.

Various blocks and the ways they should be averaged over disorder are shown in Fig.\ \ref{fig:Blocks-MT}. As usual, the basic element is the single Green's function averaged in the non-crossing approximation: $G^{R,A}(E,p) = 1/[E-\xi\pm i/2\tau]$, where $\xi(p)=p^2/2m-\mu$ \cite{AGD}. Then one should draw all possible impurity lines without intersections, that results in a number of impurity ladders (generalized cooperons) \cite{Akkermans}.

Consider, for example, the averaging of a typical block of Fig.~\ref{fig:Blocks-MT}(f), which is reponsible for the AMT contribution. The energy and momentum arguments of the Green's functions are indicated in Fig.~\ref{fig:Block-AMT}. Assuming the momentum $q$ carried by the fluctuation propagator satisfies $q\ll p_F$, we get for the corresponding block: 
\begin{equation}
{\cal{B}}^\text{MT}_\text{(f)}
=
-
\int (d\mathbf{p}) v_x^2
\left<
G^R_EG^A_{\Omega-E}G^A_{E-\omega}G^R_{\Omega+\omega-E}
\right> ,
\end{equation}
where $v_x$ is the $x$-component of the velocity, and
the sign is due to its opposite directions at the left and right current vertices.
After averaging one obtains two generalized cooperons: $\gamma^A_{2E-2\omega-\Omega}(q)$ and $\gamma^A_{-2E+\Omega}(q)$ in the lower and upper corners, respectively. Integrating first over $\xi$ and then averaging over the directions of momentum at the Fermi surface with the help of Eq.\ \eqref{f-def}, one gets:
\begin{multline}
\label{eq:Bf}
{\cal{B}}^\text{MT}_\text{(f)}(E,\Omega,\omega,q)
=-
2\pi\nu D\tau^2
\gamma^A_{2E-2\omega-\Omega}(q)\gamma^A_{-2E+\Omega}(q)
\\
\times
\frac{
f_1^A(2E-2\omega-\Omega,q)+f_1^A(-2E+\Omega,q)}{(1-i\omega\tau)^2}
.
\end{multline}
The part with the propagator and distribution functions corresponding to the block of Fig.\ \ref{fig:Blocks-MT}(f) has the form:
\begin{equation}
\label{eq:Lf}
{\cal{L}}^\text{MT}_\text{(f)}
=
\left[F_{E-\Omega}-F_{E-\Omega-\omega}\right]
\left[F_EL^A_\Omega+L^K_\Omega-F_{E-\omega}L^R_\Omega\right].
\end{equation}
Finally, the AMT contribution $Q_\text{(f)}^\text{MT}(\omega)$ is determined by Eqs.\ \eqref{eq:Qf}, \eqref{eq:Bf} and \eqref{eq:Lf}.

This procedure should be repeated with other blocks of $G^R$ and $G^A$, and then with others diagrams, leading to the final expression for the kernel $Q(\omega)$. To find the static conductivity, we extract the linear in $\omega$ term from $Q(\omega)$. After that we simplify the general expression by shifting the variable $E\to E-\Omega$ in a number of terms and integrating by parts over $E$. Additional simplification is achieved by using the detailed balance identity 
\begin{equation}
B_\Omega(F_E-F_{E-\Omega})=1-F_EF_{E-\Omega}
\end{equation}
and its derivatives.

\section{AL contribution and its decomposition into AL1 and AL2}
\label{AA:AL}

Staring with Eq.~\eqref{MAS-QAL-ini} and following the approach described in Appendices \ref{A:Initial expression} and \ref{A:Disorder averaging}, after some algebra we obtain the following expression for the AL correction:
\begin{multline}
\label{AL-ordinary}
\delta\sigma^\text{AL}
=
2\pi^{2}D\tau^{3}\int(d\Omega)(dq)
\\{}
\times
\Big\{
B^{\prime}_\Omega
\left[
L^R L^A(\Phi^R+\Phi^A)^2-2(L^R\Phi^R+L^A\Phi^A)^2
\right]
\\{}
+2B_\Omega
\left[
(L^R)^2\Phi^R\Psi^R+(L^A)^2\Phi^A\Psi^A
\right]
\Big\} ,
\end{multline}
where the arguments of the functions in the square brackets are $\Omega$ and $q$.
In Eq.~(\ref{AL-ordinary}), the functions $\Phi$ and $\Psi$ originating from averaging of blocks with three Green's functions are given by
\begin{subequations}
\label{PhiPsi}
\begin{gather}
\label{Phi-def}
\Phi^{R}(\Omega,q)
=
\int(dE)F_Eg_2^{A}(\gamma^{A})^2 ,
\\
\label{Psi-def}
\Psi^{R}(\Omega,q)
=\int(dE)F^{\prime}_E\left[
2g_1^{A}\gamma^{A}+g_2^{A}(\gamma^{A})^2
\right].
\end{gather}
\end{subequations}
where the arguments of the functions $g_m$ and $\gamma$ are $2E-\Omega$ and $q$.

The AL contribution \eqref{AL-ordinary} can be naturally decomposed into a sum of the terms with one (AL1) and two (AL2) fluctuation propagators. To this end we note that the function $\Phi^R$ can be represented as the derivative of the inverse fluctuation propagator:
\begin{equation}
\label{L^2-L}
\Phi^{R}(\Omega,q)= -\frac{i}{2 \pi\tau l} 
\frac{\partial [L^R(\Omega,q)]^{-1}}{\partial q} .
\end{equation}
Therefore the term proportional to $B_\Omega$ in Eq.\ (\ref{AL-ordinary}) can be expressed via the derivative of a single propagator:
\begin{equation}
\label{AL1-befor}
\delta\sigma^\text{AL1}
=
-\frac{4\pi D\tau^{2}}{l} 
\int(d\Omega) (d\textbf{q}) B_\Omega
\Im(\Psi^R\partial_q L^R) .
\end{equation}
Then integrating by parts over $|q|$ we transform $\delta\sigma^\text{AL1}$ to the form of Eqs.~(\ref{ds-via-Sigma}) and (\ref{S_R-AL1}). Finally, with the help of Eq.\ \eqref{L^2-L} the remaining part with two propagators, $\delta\sigma^\text{AL2}=\delta\sigma^\text{AL}-\delta\sigma^\text{AL1}$, can be brought to the form (\ref{sigma-AL2-v2}) which contains only fluctuation propagators.

\section{Equivalence to Ref.~\cite{Tikhonov} in the diffusive region}
\label{A:Comp2Kostya}

In this Appendix we demonstrate that our expression for the fluctuation correction coincides with the result of Ref.~\cite{Tikhonov} at zero magnetic field. We start with rewriting $\delta\sigma^\text{(dos)}$ and $\delta\sigma^\text{(sc)}$ given by Eqs.~(77) and (79) of Ref.~\cite{Tikhonov} in the representation of Eq.~(\ref{sigma-varsigma}). Taking the limit of zero magnetic field and expressing everything in terms of the function $G(z)$ [see Eq.~(\ref{G(z)})], we obtain:
\be
\label{Tikhonov-def-dos}
\varsigma^\text{(dos)}
=-b
\Im\frac{G''}{G}
-b'
\frac{\Im G\Im G'}{|G|^{2}} ,
\ee
\\[-30pt]
\begin{multline}
\label{Tikhonov-def-sc}
\varsigma^\text{(sc)}
=
b x \Im
\frac{G'G''}{G^{2}}
\\{}
+
b'x
\left(
4 \frac{\Re G'\Im G}{|G|^{2}} \Im \frac{G'}{G} 
- \Im G'\Im\frac{G'}{G^{2}}\right) .
\end{multline}

In order to compare $\delta\sigma^\text{(dos)}+\delta\sigma^\text{(sc)}$ with $\delta\sigma^\text{REG}+\delta\sigma^\text{AL2}$,
we first write $G'/G^2$ in the first term in Eq.~(\ref{Tikhonov-def-sc}) as $-\partial G^{-1}/\partial x$ and integrate it over $x$ by parts. 
This procedure leads to the cancellation of the linear in $b$ term in the difference
\begin{equation}
\delta\varsigma
=
\varsigma^\text{REG}+\varsigma^\text{AL2}
-\varsigma^\text{(dos)}-\varsigma^\text{(sc)} ,
\end{equation}
where integration over $x$ according to Eq.~(\ref{sigma-varsigma}) is implied.
Then we single out the terms in $\delta\varsigma$ which do not contain the factor of $x$ (originating from $\delta\varsigma^\text{REG}-\delta\varsigma^\text{(dos)}$), write there $G'+xG''=\partial (xG')/\partial x$, and integrate over $x$ by parts.
Using the identity
\begin{equation}
\Im \frac{G'}{|G|^2}+\Im \frac{G'}{G^2}
=
2\Re\frac{G}{|G|^2}\Im\frac{G'}{G},
\end{equation}
we reduce the difference to the form
\begin{equation}
\delta\varsigma 
=4 b' x
\Im\frac{G'}{G}\frac{\Im G'\Re G-\Re G'\Im G-\Im G'G^*}{|G|^2} ,
\end{equation}
which equals zero due to a general identity
$\Im A\Re B-\Re A\Im B=\Im AB^*$.

Finally, one can easily verify that 
$\delta\sigma^\text{(an)}=\delta\sigma^\text{AMT}$,
proving full equivalence of our approach to that of Ref.~\cite{Tikhonov}.

\section{Anomalous MT correction in the diffusive region near the transition}
\label{A:AMT}

In this Appendix, we compute the expansion of $\delta\sigma^\text{AMT}$ in the vicinity of the transition, $\epsilon\ll1$. The dephasing rate is also assumed to be small, $\gamma\ll1$, but the relation between $\epsilon$ and $\gamma$ can be arbitrary.

The most singular contribution \cite{MT1,MT2} originates from small $x$ and $y$, where one can replace $G$ by $G_0=(\pi^2/2)(x+iy)+\eps$ and $b(y)$ by the leading asymptotics $b_0(y)=1/(2\pi y)$. Then substituting Eq.~(\ref{varsigma-AMT}) into Eq.~(\ref{sigma-varsigma}) and integrating over $y$, we arrive at
\be
\label{sigma-varsigma-AMT-0}
\delta\sigma^\text{AMT}_0
=
\frac{1}{4\pi^2}
\int_0^\infty dx
\frac{1}{x+x_*}
\frac{1}{x+x_\epsilon} ,
\ee
where $x_\epsilon = (2/\pi^2)\eps$.
Thus one readily recovers the leading asymptotics given by the first term in Eq.~(\ref{sigma-AMT-res-diff-close}).

The terms omitted in the derivation of Eq.~(\ref{sigma-varsigma-AMT-0}) can be written as
\be
\label{delta-sigma-AMT-1}
\delta\sigma^\text{AMT}_1
=
-
\frac{1}{2\pi^2}
\int_0^\infty
\frac{dx}{x+x_*}
\int_{-\infty}^\infty dy \,
H(x,y,\epsilon)
,
\ee
where 
\be
\label{H(x,y,e)}
H(x,y,\epsilon)
=
b'(y)
\frac{\Im^2 G(z)}{|G(z)|^2} 
-
b'_0(y) \frac{\Im^2 G_0(z)}{|G_0(z)|^2}
.
\ee
In order to extract the leading contribution from Eq.~(\ref{delta-sigma-AMT-1}) at small $\epsilon$, one has to study the behavior of the function $H(x,y,\epsilon)$ in the limit $x,\eps\to 0$. This should be done with care, since the resulting function contains both a smooth part decaying at $y\sim1$, and a sharp $\delta$-function like peak originating from the first term in Eq.~(\ref{H(x,y,e)}):
\begin{multline}
\label{H-lim}
H(x,y,\epsilon)
\approx
b'(y) \frac{\Im^2 \psi(1/2+iy)}{|\psi(1/2+iy)-\psi(1/2)|^2} - b'_0(y)
\\{}
+ 
\frac{14 \zeta (3)}{\pi^4}
\frac{(x+x_e) (x^2+2 x x_e+y^2)}{[(x+x_e)^2+y^2]^2} .
\end{multline}
Substititing Eq.~(\ref{H-lim}) into Eq.~(\ref{delta-sigma-AMT-1}), we arrive at 
\begin{multline}
\label{delta-sigma-AMT-1-res}
\delta\sigma^\text{AMT}_1
=
-
\left[ \frac{7 \zeta (3)}{\pi^4} + \kappa \right]
\ln\frac{1}{\gamma}
\\{}
+
\frac{7\zeta(3)}{2 \pi ^4}
\frac{\epsilon [\epsilon \ln(\epsilon/\gamma) - \epsilon + \gamma]}
{(\epsilon -\gamma )^2}
+ {\cal{O}}(1) ,
\end{multline}
where $\kappa=0.1120$ is the value of the integral
\be
\label{kappa}
\kappa 
=
\frac{1}{2\pi^3}
\int_{0}^\infty dy
\left[
\frac{1}{y^2}
- 
\frac{[\psi'(1/2)]^2\mathop{\rm sech}^4(\pi y)}{|\psi(1/2+iy)-\psi(1/2)|^2}
\right] .
\ee

\section{$L^R$, $\Sigma^R$, and $\Sigma^Z$ in the clean limit}
\label{A:Sigma-bal}

In this Appendix we refine the basic ingredients of Eq.\ \eqref{ds-via-Sigma-ball}, $L^R$, $\Sigma^R$, and $\Sigma^Z$, in the ballistic limit.
For the functions $\Sigma^R$ and $\Sigma^Z$ we use Eqs.\ \eqref{SR-clean} and \eqref{SZ-clean} which provide the leading asymptotic behavior at $\tau\to\infty$.

\subsubsection{\textbf{Fluctuation propagator}}

In the ballistic limit, the function $f^R_1$ defined in Eq.\ \eqref{f-def} scales as $1/\tau$, and the fluctuation propagator \eqref{LR-def} becomes $\tau$-independent. Summation over the Matsubara energies can then be easily performed, leading to
\be
\label{LR-inv-ball}
(L^R)^{-1}
=
\epsilon + \corr{\psi(1/2+ix\cos\theta-iy)} - \psi(1/2) ,
\ee
where the dimensionless momentum, $x$, and frequency, $y$, in the ballistic region are defined in Eq.\ \eqref{xy-ball},
and $\corr{\cdots} = \int (\cdots) d\theta/2\pi$ stands for the angular averaging in 2D. Equation \eqref{LR-inv-ball} can be brought to a form more suitable for further evaluation with the help of the integral representation for the digamma function \cite{GR}, which we present in the form with 1/2 isolated:
\be
\label{psi-integral}
\psi(1/2+z)
=
\int_0^\infty 
\left[
\frac{e^{-t}}{t}
-
\frac{e^{-zt}}{2\sinh t/2}
\right] dt .
\ee
Then the angular averaging can be easily performed and we obtain
\be
\label{LR-inv-ball-res}
(L^R)^{-1}
=
\epsilon + 
\int_0^\infty 
\frac{1-J_0(xt)e^{iyt}}{2\sinh t/2}
dt ,
\ee
where $J_0(z)$ is the Bessel function.

\subsubsection{\textbf{Function $\Sigma^R$}}

In the leading order in $\tau\to\infty$, $\Sigma^R$ is given by Eq.~(\ref{SR-clean}).
We write it as $\Sigma^{R(11)} + \Sigma^{R(2)}$, in accordance with the two summands in Eq.~(\ref{SR-clean}).
In calculating $\Sigma^{R(11)}$, we use the explicit form of $f_1^A(2E-\Omega,q)=f_1^R(\Omega-2E,q)$ given by Eq.~(\ref{f1-2D-3D}),
deform the integration contour to the lower half-plane picking the poles of $F_E$, and keep the leading order in $1/\tau$:
\begin{multline}
\label{SR-clean-11}
\Sigma^{R(11)}
=
- 
\frac{4iT}{\tau^2}
\sum_{\eps>0}
\frac{\partial}{\partial E}
\frac{1}{(2iE-i\Omega)^2+q^2v^2}
\biggr|_{E=-i\eps} 
\\{}
=
-i
\frac{\psi'(1/2+ix-iy)-\psi'(1/2-ix-iy)}{16 \pi^3 x T^2 \tau^2} .
\end{multline}
On the other hand, in order to calculate $\Sigma^{R(2)}$ we find it more convenient to keep $f_2$ in the original form (\ref{f-def}), that allows to perform integration over $E$ similar to Eq.~(\ref{SR-clean-11}):
\begin{multline}
\label{SR-clean-2}
\Sigma^{R(2)}
=
\frac{4iT}{\tau^2}
\sum_{\eps>0}
\frac{\partial}{\partial E}
\left< \frac{1}{(2iE-i\Omega+iqv\cos\theta)^2} \right> 
\biggr|_{E=-i\eps} 
\\{}
=
- \frac{\corr{\psi''(1/2+ix\cos\theta-iy)}}{8 \pi^3 T^2\tau^2}
,
\end{multline}
where the angular averaging is still to be done.

Since $\Sigma^R = \Sigma^{R(11)} + \Sigma^{R(2)}$, it is convenient to bring Eqs.~(\ref{SR-clean-11}) and (\ref{SR-clean-2}) to a similar form. That can be done with the help of Eq.~\eqref{psi-integral}, leading to
\be
\label{SR-clean-res}
\Sigma^{R}
=
\frac{1}{8 \pi^3 T^2 \tau^2} 
\int_0^\infty 
\frac{t^2 [J_0(xt) - S_0(xt)] e^{iyt}}{2\sinh t/2}
\, dt ,
\ee
where $S_0(z)\equiv \sin(z)/z$.
The terms with $J_0$ and $S_0$ correspond to $\Sigma^{R(2)}$ and $\Sigma^{R(11)}$, respectively.

\subsubsection{\textbf{Function $\Sigma^Z$}}

In the leading order in $\tau\to\infty$, $\Sigma^Z$ is given by Eq.~(\ref{SZ-clean}). In this approximation we neglect cooperons and therefore a special region of low momenta, $ql\ll1$, shown in Fig.~\ref{F:cuts}(a) does not appear. Following Eq.~(\ref{SZ-clean}), we write $\Sigma^Z$ as $\Sigma^{Z(11)} + \Sigma^{Z(2)}$. The first term, $\Sigma^{Z(11)}$, involves the product of $f_1^Rf_1^A$, which has branch cuts both in the upper and lower half-planes of complex $E$, as shown in Fig.~\ref{F:cuts}(b). 
Therefore it can be written as the sum of the contributions from the poles, $\Sigma^{Z(11)}_\text{tanh}$, and from the cut, $\Sigma^{Z}_\text{sing}$. The former is calculated similar to Eq.~(\ref{SR-clean-11}):
\begin{multline}
\label{SZ-clean-11-poles}
\Sigma^{Z(11)}_\text{tanh}
=
\Re \frac{8iT}{\tau^2}
\sum_{\eps>0}
\frac{1}{(2iE-i\Omega)^2+q^2v^2}
\biggr|_{E=-i\eps} 
\\{}
=
\frac{\Re[\psi(1/2+ix-iy)-\psi(1/2-ix-iy)]}{4 \pi^2 x T \tau^2} .
\end{multline}
The contribution from the branch cut contains an additional logarithmic singularity in the limit $T\tau\to\infty$, originating from the pinching of the cuts in the upper and lower half-planes [see Fig.~\ref{F:cuts}(b)]. Making the energy shift, $E=E'+\Omega/2$, we obtain:
\be
\Sigma^{Z}_\text{sing}
\approx
- \frac{8}{\tau^2}
\int_{-qv/2+1/\tau}^{qv/2-1/\tau} \frac{dE}{2\pi}
\frac{F(E'+\Omega/2)}{q^2v^2-(2E')^2} .
\ee
The logarithmic divergency is regularized at $|E'|-qv/2\sim1/\tau$, which is the vertical distance between the two branch cuts.
With logarithmic accuracy we obtain
\begin{equation}
\label{SZ-clean-sing-res}
\Sigma^{Z}_\text{sing}
\approx
-
\frac{\tanh\pi(x+y)-\tanh\pi(x-y)}{4\pi^2T\tau^2x} \ln T \tau x .
\end{equation}

Calculation of $\Sigma^{Z(2)}$ is completely analogous to that of $\Sigma^{R(2)}$ in Eq.~(\ref{SR-clean-2}):
\begin{multline}
\label{SZ-clean-2}
\Sigma^{Z(2)}
=
\Re \frac{8iT}{\tau^2} \sum_{\eps>0}
\left< \frac{1}{(2iE-i\Omega+iqv\cos\theta)^2} \right> 
\biggr|_{E=-i\eps} 
\\{}
=
\frac{\Re i \corr{\psi'(1/2+i x \cos\theta-iy)}}{2 \pi ^2 T\tau^2} .
\end{multline}

Now using Eq.~(\ref{psi-integral}) we reduce the sum of Eqs.~(\ref{SZ-clean-11-poles}) and (\ref{SZ-clean-2}) to a concise form:
\be
\label{SZ-clean-poles-res}
\Sigma^{Z}_\text{tanh}
=
- \frac{1}{2 \pi^2 T \tau^2} 
\int_0^\infty 
\frac{t[J_0(xt)+S_0(xt)] \sin(yt)}{2\sinh t/2} dt
.
\ee
The resulting expression for $\Sigma^Z=\Sigma^Z_\text{tanh}+\Sigma^Z_\text{sing}$ is the sum of the two contributions given by Eqs.~\eqref{SZ-clean-poles-res} and \eqref{SZ-clean-sing-res}, respectively.

\subsubsection{\textbf{Coefficients in Eqs.~(\ref{ds-ball-Z-poles}) and (\ref{ds-ball-Z-sing})}}

The coefficients $\alpha_p$ and $\alpha_c$ in Eqs.~(\ref{ds-ball-Z-poles}) and (\ref{ds-ball-Z-sing}) are given by the following integrals:
\begin{multline}
\label{int-ap}
\alpha_p
=
\int_0^\infty x \, dx
\int_{-\infty}^\infty dy
\int_0^\infty ds
\int_0^\infty t\, dt
\\{}
\times
\frac{J_0(xs) [J_0(xt)+S_0(xt)] \sin(ys) \sin(yt)}
{\sinh(s/2)\sinh(t/2) \sinh^2(2\pi y)} 
,
\end{multline}
\vskip-1mm
\begin{multline}
\label{int-ac}
\alpha_c 
=
\int_{-\infty}^\infty dy
\int_0^\infty dx
\int_0^\infty dt \,
\frac{J_0(xt)\sin(yt)}{\sinh t/2}
\\{}
\times
\frac{\tanh\pi(x+y)-\tanh\pi(x-y)}{\sinh^2(2\pi y)} 
.
\end{multline}

\end{document}